\documentclass{article}

\usepackage{arxiv}

\usepackage[utf8]{inputenc} 
\usepackage[T1]{fontenc}    
\usepackage{hyperref}       
\usepackage{url}            
\usepackage{booktabs}       
\usepackage{amsfonts}       
\usepackage{nicefrac}       
\usepackage{microtype}      
\usepackage{lipsum}
\usepackage{graphicx,subcaption}
\usepackage{amsmath}
\usepackage{amssymb}

\usepackage{graphicx}%
\usepackage{multirow}%
\usepackage{amsmath,amssymb,amsfonts}%
\usepackage{amsthm}%
\usepackage{mathrsfs}%
\usepackage[title]{appendix}%
\usepackage{xcolor}%
\usepackage{textcomp}%
\usepackage{manyfoot}%
\usepackage{booktabs}%
\usepackage{algorithmicx}%
\usepackage{algpseudocode}%
\usepackage{listings}%

\usepackage{bbm}
\usepackage{caption}
\usepackage{subcaption}
\usepackage[linesnumbered]{algorithm2e}
\usepackage{multirow}

\newcommand{\bx}{\mathbf{x}}

\def\bY{\boldsymbol{Y}}

\def\bx{\boldsymbol{x}}

\theoremstyle{definition}
\newtheorem{Example}{Example}[section]
\newtheorem{Remark}{Remark}[section]

\theoremstyle{thmstyleone}%
\theoremstyle{thmstyletwo}%

\theoremstyle{thmstylethree}%

\DeclareMathOperator*{\argmax}{arg\,max}
\def\bY{\mathbf{Y}}

\title{A Novel Criterion for Interpreting Acoustic Emission Damage Signals Based on Cluster Onset Distribution}
\date{}

\begin{document}
\maketitle

\vspace{-2cm}

\begin{center}
  \bf Emmanuel Ramasso$^{1*}$,  Martin Mbarga Nkogo$^{1}$, Neha Chandarana$^{2}$, Gilles Bourbon$^{1}$, Patrice Le Moal$^{1}$, Quentin Lefèbvre$^{1}$, Martial Personeni$^{1}$, Constantinos Soutis$^{3}$, Matthieu Gresil$^{4}$, Sébastien Thibaud$^{1}$ 
\end{center}

\vspace{0.5cm}

\noindent $^{1*}$Université Marie et Louis Pasteur, CNRS, UTBM, institut FEMTO-ST, F-25000, Besan\c con, France\\
\noindent $^{2}$Bristol University, Department of Aerospace Engineering, United Kingdom\\
\noindent $^{3}$Department of Materials, The University of Manchester, Manchester, UK \&
Aerospace Research Institute, The University of Manchester, Manchester, UK\\
\noindent $^{4}$i-Composites Lab, Department of Materials Science and Engineering \& Department of Mechanical and Aerospace Engineering, Monash University, Clayton, VIC, Australia\\
\noindent $^{*}$\texttt{emmanuel.ramasso@femto-st.fr}  

\vspace{2cm}

\begin{abstract}
Structural health monitoring (SHM) relies on non-destructive techniques such as acoustic emission (AE) that generate large amounts of data over the lifespan of systems. Clustering methods are used to interpret these data and gain insights into damage progression and mechanisms. Conventional methods for evaluating clustering results utilise clustering validity indices (CVI) that prioritise compact and separable clusters. This paper introduces a novel approach based on the temporal sequence of cluster onsets, indicating the initial appearance of potential damage and allowing for early detection of defect initiation. The proposed CVI is based on the Kullback-Leibler divergence and can incorporate prior information about damage onsets when available.

Three experiments on real-world datasets validate the effectiveness of the proposed method. The first benchmark focuses on detecting the loosening of bolted plates under vibration, where the onset-based CVI outperforms the conventional approach in both cluster quality and the accuracy of bolt loosening detection. The results demonstrate not only superior cluster quality but also unmatched precision in identifying cluster onsets, whether during uniform or accelerated damage growth. The two additional applications stem from industrial contexts. The first focuses on micro-drilling of hard materials using electrical discharge machining, demonstrating, for the first time, that the proposed criterion can effectively retrieve electrode progression to the reference depth, thus validating the setting of the machine to ensure structural integrity. The final application involves damage understanding in a composite/metal hybrid joint structure, where the cluster timeline is used to establish a scenario leading to critical failure due to slippage.
\end{abstract}

\keywords{Clustering Validity index\and Acoustic Emission\and Clusters timeline\and Structural Health Monitoring}

\section{Introduction}

{{

Ensuring the availability of equipment and the integrity of structures is a critical concern within the aeronautical and civil engineering industries, both economically and for user safety. This necessitates the monitoring of structures as well as understanding potential damage mechanisms that can indicate the integrity of the structure. Structural health monitoring (SHM) is a research and development field aimed at implementing and deploying solutions to monitor engineering systems such as equipment and structures \cite{FarrarBook,giurgiutiu2007structural,Mardanshahi25}.

During operation, systems and structures are subjected to mechanical, electrical, or chemical stresses, which generate localised micro-deformations. These micro-deformations may be accompanied by the release of energy in the form of elastic waves, known as acoustic emission (AE). When these changes are irreversible and likely to degrade performance, they are referred to as damage. Accumulated micro-damages eventually lead to material failure, rendering the structure unusable, and potentially leading to catastrophic failure \cite{manuello2020acoustic}. Piezoelectric transducers can be placed onto or embedded in the host structure to record and convert these micro-vibrations into usable electrical signals \cite{giurgiutiu2007structural,NehaPhD}. The recorded signals contain the signatures of the types of damage from which they have been generated \cite{zhang2024acoustic,Holford2017,al2015classification,melchiorre2023acoustic}. 

As a passive technique, it exploits the transient signals emitted during damage occurrence. Therefore, sensors must be continuously monitored or triggered by significant damage. Although the sampling frequency is relatively high (typically 5 to 10 MHz, depending on the sensors), it is possible to detect damage in real time from AE data streams, even during continuous monitoring, thanks to modern computing capabilities. A key challenge in real-time processing is the acoustic emissivity of the monitored material and/or structure; the more emissive the material and/or structure, the greater the number of AE sources it has, leading to more signals that need to be processed, which increases computational costs. To manage this, an amplitude threshold is typically applied to limit the number of transients to be analysed. After distinguishing between noise and meaningful data, various methodologies can be applied to interpret AE data streams. Generally, these methodologies share a common focus on analysing certain statistical properties, although the methods for deriving these properties vary significantly across publications. The relevance of the AE technique has been demonstrated in various applications for decades, in particular for tool wear condition monitoring during in milling, grinding and drilling, as well as for characterisation and monitoring of various materials, such as composite materials and rocks \cite{Chen2025,Alsuhaibani2025,Tai2025}. 

\medskip

Correctly interpreting acoustic emission data is of key importance for predictive maintenance in machining processes \cite{Ferrisi2025}. However, due to the lack of knowledge concerning the precise source of each AE signal, the interpretation of acoustic emission data remains an open issue. 

One approach is \textit{sample-based analysis}. Samples can be analyzed \textit{point by point} \cite{pomponi2013real,placet2014online} for wave picking and clustering. Another way is by a \textit{sliding window}. For example, Martin-Del-Campo \textit{et al.} \cite{martin2019detection,martin2017online} developed an anomaly detector where part of the data is labelled as "normal", and the algorithm detects deviations, providing insights into structural integrity of bearings with particle-contaminated lubricant at varying rotational speeds. The method relies on dictionary learning using waveforms matched to a windowed portion of the stream, optimised to produce a sparse and accurate code, thus reducing the measurement data rate by an order of magnitude. 

Data analysis can also be performed \textit{at the transient level}. Transient signals must first be detected using wave selection methods from the literature \cite{kurz2005strategies,Pomponi2015110,MADARSHAHIAN2019483,QUY2021107254,Holford2017,kharrat2016signal}. Interpretation follows and aims to address challenges such as damage localisation or clustering for damage classification. For example, Chen \textit{et al.} \cite{chen2017similarity} applied dynamic time warping to measure the similarity of AE signals for source localisation when multiple AE sources are present. Data from thermal-cracking experiments on granite were used to validate the approach. In \cite{ramasso2020learning}, randomly selected signals are represented by auto-regressive hidden Markov models, subsequently applied to all transients of a dataset to obtain scores indicating goodness-of-fit, which are then used as inputs for clustering algorithms. This method was validated on datasets from carbon fibre composites.

\medskip

One of the most frequently applied methodologies encountered in the literature relies \textit{on three main steps} (Figure \ref{fig:zfezefzfz}): 1) hit detection - where transient signals are detected based on signal processing tools - and feature extraction where the transient signals are compressed and represented in a common high dimensional space for further analysis, 2) clustering or classification, which either groups features in an unsupervised manner or uses {{prior}} labels in a supervised approach to predict signal classes, and 3) evaluation and validation of the groups or classes of signals. 

\begin{figure}[hbtp]
\centering
\includegraphics[angle=0,width=0.99\columnwidth]{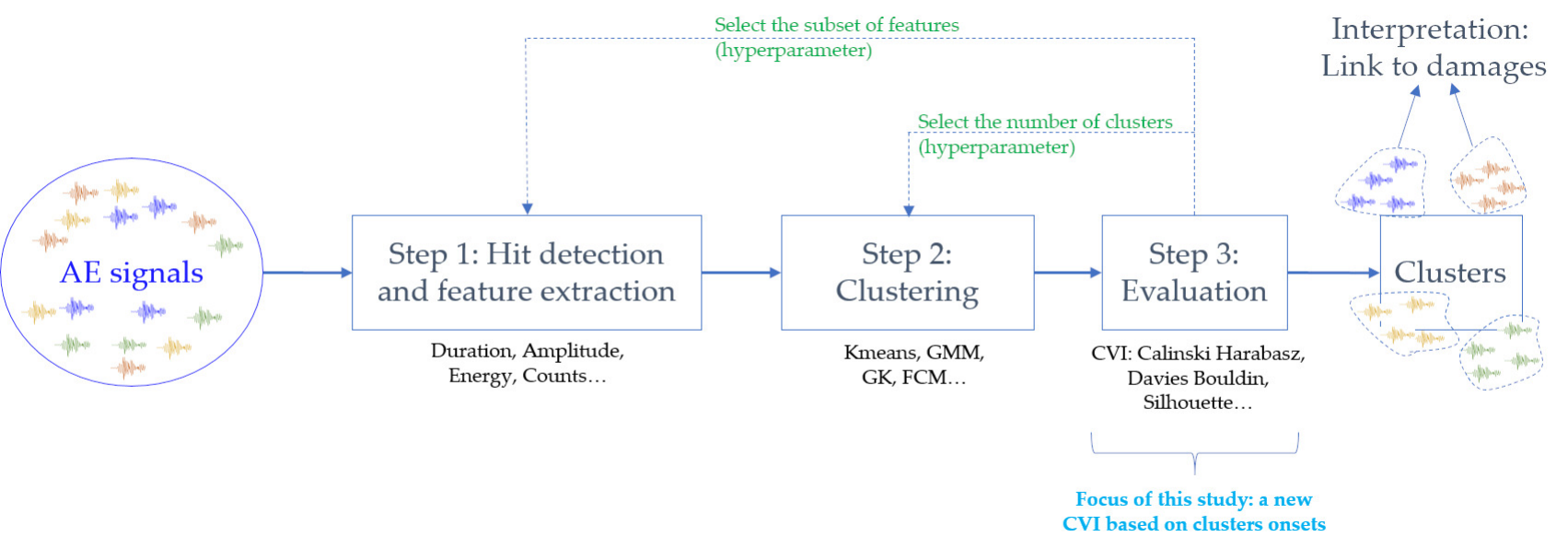}
\caption{Feature-based clustering of AE data.}
\label{fig:zfezefzfz}
\end{figure}

This methodology has been applied, for example, to carbon fibre composite panels for aeronautics \cite{al2015classification}, high-performance carbon fibre-reinforced thermoset composites in extreme environments \cite{ramasso2015unsupervised}, bridge and tunnel health assessment \cite{li2024deep}, railway systems \cite{carboni2020acoustic}, tribology \cite{olorunlambe2022bio,tervo2024hybrid}, and fatigue crack monitoring \cite{karimian2020new}. Deep learning approaches, which are supervised methods, follow similar steps but typically automate feature extraction through layers during training \cite{du2024using,hesser2021identification,ramasso2024condition}.

\medskip

The final step, namely \textit{validation}, \textit{which is central to this work}, is more challenging for clustering than for supervised learning due to its subjectivity \cite{carmichael1968finding}, especially given the lack of precise ground truth in the interpretation of AE data. Various criteria, known as \textit{clustering validity indices} (CVI), have been proposed in the pattern recognition community \cite{arbelaitz2013extensive,maulik2002performance,vendramin2010relative,vendramin2013combination,bezdek1998some,baya2023dstab} to determine whether clusters are meaningful.

For a given application, the challenge for unsupervised interpretation of AE data is the extraction of relevant features that can link the identified clusters to damage mechanisms. This involves answering the following key questions: \textit{What are the most relevant features? How many clusters are needed?}

An answer to these questions is provided by an algorithm used in the literature detailed in appendix \ref{app1} and proposed by \cite{sause12}. It uses several CVIs evaluated for different combinations of $k\cdot C_n^p$ (subsets of $p$ features from $n$, considering $k$ different cluster counts). A majority voting scheme, weighted by the user, is used to select the best number of clusters and subset of features. 

The CVI used for AE data interpretation, such as the Davies-Bouldin index \cite{davies1979cluster} or Silhouette index \cite{rousseeuw1987silhouettes}, \textit{are exclusively focused on clusters' shape, characterised by their compactness and separability in the feature space}. One problem with these CVIs is that the chosen metric (such as Euclidean distance) may differ from those used in the clustering method's objective function (such as likelihood), potentially leading to misleading interpretations \cite{sawan2015unsupervised,ramasso2015unsupervised}. Sawan \textit{et al.} \cite{sawan2015unsupervised} showed that acoustic emission data should be divided into more clusters than are expected from the number of active damage mechanisms to prevent misrepresentative grouping. Similarly, the results of several clustering methods were combined in \cite{ramasso2015unsupervised} to more accurately map the feature space, resulting in a timeline that reflects a plausible damage scenario. Another issue with classical CVIs is that the timeline provided by compact and separated clusters is generally not related to the expected damage scenario, hence justifying novel approaches to achieve more relevant clustering performance \cite{ramasso2014reconnaissance,sawan2015unsupervised,chandarana2018damage,hohl2018computationally,rastegaev2018using,NehaPhD,Vinogradov20}. These works show that damage can be described by a set of clusters, which may or may not correspond to true mechanisms, but the common factor is that one cluster should act as a precursor to others, following a progression. 

In previous literature, clustering results are often presented as sequences after applying shape-based CVIs. However, current methods do not assure users that the results will be relevant in terms of the timeline. Therefore, the challenge remains: how can we select features and determine the number of clusters to guarantee a relevant timeline?

}}

\medskip

\begin{figure}[hbtp]
\centering
\subfloat[\centering Selection" (left) is based on clusters shape in the feature space in low-dimensional space for visualization purpose. "Representation" (right) of clusters of the timeline provides onsets which start too early.]{\includegraphics[width=0.75\columnwidth]{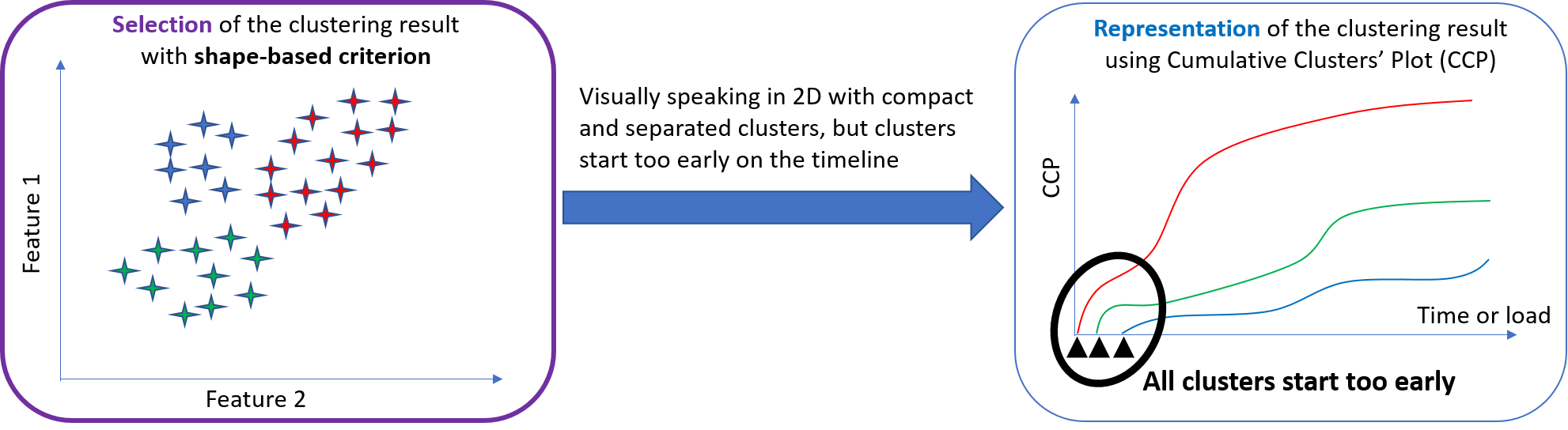}}  \\
\subfloat[\centering "Selection" (right) is based on clusters onsets. The timeline is relevant for monitoring. The "representation" (left) in low-dimension (2D in general) of the clusters can be useless because the clusters are compact and separated only in higher dimensions (used in clustering). The compactness and separation of clusters is ensured during clustering.]{\includegraphics[width=0.75\columnwidth]{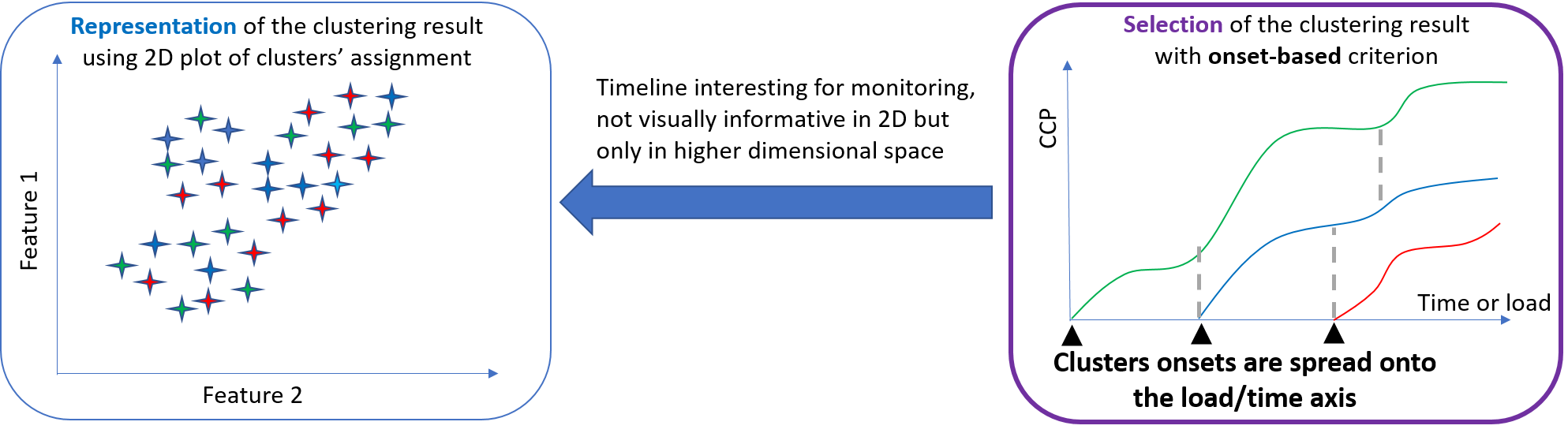}} 
\caption{Selection and representation of clusters results. \label{fig:quality}}
\end{figure}

{{

Our work aims to address this question by introducing a new validation criterion based on cluster onsets. The main features and advantages of this criterion are as follows:

\begin{itemize}

\item The proposed criterion is the first to formalise the ranking of cluster partitions based on how clusters emerge over time or in relation to a physical parameter. This approach provides a dynamic perspective on clustering behaviour, offering a new way to assess the evolution of clusters beyond static measures. The criterion is innovative and lacks a direct equivalent in the existing literature for selecting the number of clusters and the feature subsets that best explain the data in terms of a timeline.  

\item The proposed criterion is valuable for onset-based interpretation of AE signals particularly in cases where materials degrade progressively under stress. Three challenging experiments on real-world datasets demonstrate the relevance of the proposed approach. One of the applications is a benchmark on the detection of the loosening of bolted plates under vibration where the proposed criterion outperforms existing methods. It provides superior results, both quantitatively (in terms of accuracy) and qualitatively (in terms of the positioning of estimated onsets and cumulative occurrences). 

\end{itemize}

To illustrate the positioning of our contribution, Figure \ref{fig:quality} highlights the fundamental \textit{difference between shape-based and onset-based approaches} for evaluating the results of AE data clustering. In the shaped-based approach (Figure \ref{fig:quality}a), clusters may appear compact and well-separated in a low-dimensional feature space. However, this does not guarantee that they will correspond to relevant onsets. For example several of them can start very early and almost simultaneously when considering the horizontal axis. In contrast, the latter approach focuses on identifying feature subsets that highlight onsets distributed along the horizontal axis (such as load or time). This is more helpful for interpreting the dynamic behaviour of the material or structure under test, as illustrated in Figure \ref{fig:quality}b.
}}

\medskip

\noindent Section \ref{sect4} introduces the proposed criterion and presents an algorithm for the implemented method, incorporating the proposed criterion for selecting the best parametrisation and the resulting optimal cluster partition. The relevance of the criterion is evaluated in Section \ref{sect5} using a benchmark obtained from a real structure.

\section{Proposed criterion: Sorting subsets of features with respect to onsets}
\label{sect4}

The AE data consist of $N$ $d$-dimensional feature vectors $\bx_n \in \Re^d$. Given a subset of features $S  \subset \Re^d$ and a number of clusters $K$, a partition $P$ is obtained by applying a clustering method. $P$ is represented by a vector of length $N$, where each element $P(n) \in \{1,2\dots K\}$ represents the cluster number assigned to each AE signal. For fuzzy clustering methods, $P(n)$ is the cluster with the largest membership degree. 

\begin{figure}[hbtp]
\centering
\includegraphics[angle=0,width=0.85\columnwidth]{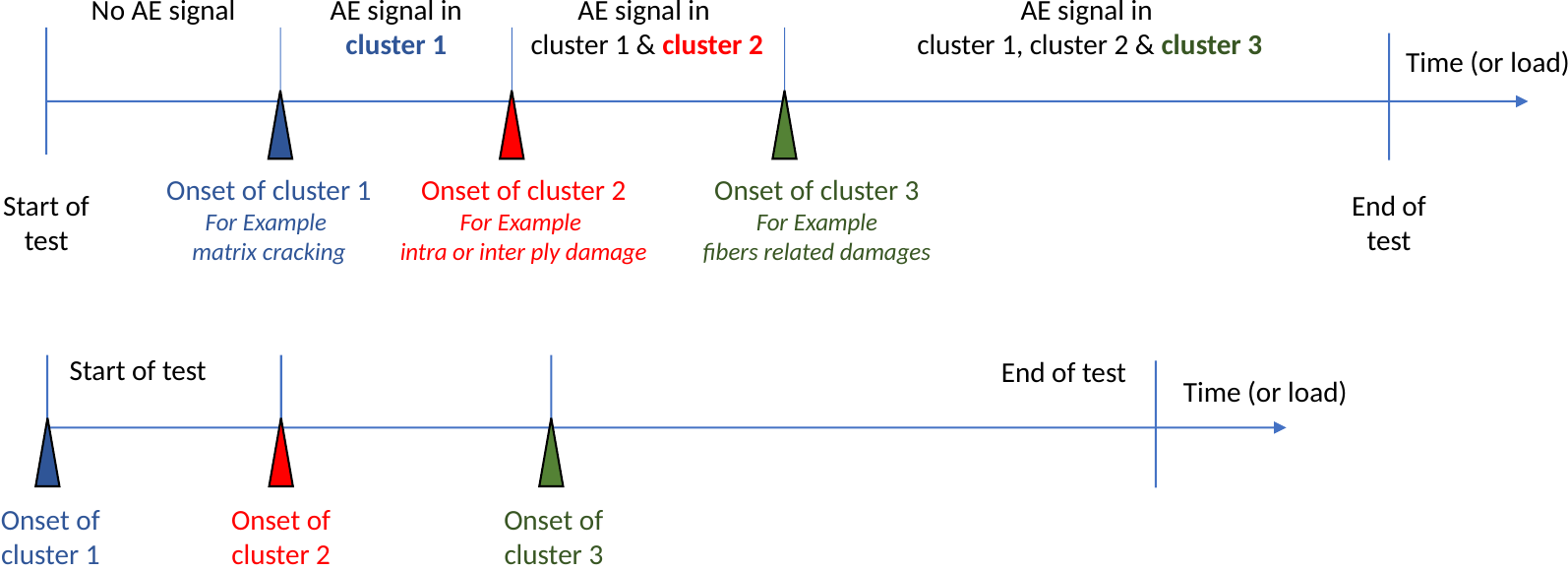} 
\caption{Illustration used in examples in Section \ref{sect4}: Onsets (triangles) are defined as the first occurrence of each cluster. The horizontal axis can be any monotonically increasing information regarding the evolution of the test like, for example, time or a damage variable. In-between onsets, several clusters, that previously occurred, can keep growing. Reference to composite damages is only for illustration. Top figure: No signal appears from the beginning to the first onset so this part is removed and the starting time shifted accordingly, which leads to the bottom figure.}
\label{fig:onsets1}
\end{figure}

\textit{A cluster's onset is defined as the first occurrence of the cluster.} For example, for cluster $k$, the onset is the first index in $P$ where the $k$-th cluster appears. Onsets can be represented on an axis related to time, load, or other physical quantities related to the degradation of the material. Illustrations can be found on the right-hand side of Figure \ref{fig:quality} and on Figure \ref{fig:onsets1}. 

Let us define $t_k\in[0,T]$ as the first occurrence of cluster $k$ in a dataset. In order to compute the criterion, the onsets are delayed so that the first one starts at $t_1=0$ (Figure \ref{fig:onsets1}). Then, we compute the normalised difference between two consecutive onsets as
\begin{equation}
\Delta_k = \frac{t_{k+1} - t_{k}}{T} \quad t_{K+1} = T
\label{eq:delta1}
\end{equation}
The set ${\mathbf{p}}_{\textrm{est}} = \{\Delta_k\}_{k=1}^K$, estimated by clustering, can be interpreted as a probability distribution made of $K$ elements since we have
\begin{equation}
\sum_{k=1}^K \Delta_k = 1
\label{eq:delta1sum}
\end{equation}
\medskip

\begin{Example}
As an example, consider Figure \ref{fig:onsets1}. We have three clusters, with onsets positioned at $t_1=27, t_2=57, t_3=82$, and with $T=127$ (horizontal axis is with arbitrary unit; in practice it can be related to time, load, or a physical quantity). We can shift the onsets to 
$t_1=0, t_2=30, t_3=55$ and $T=100$. Then $\Delta_1=30/T$, $\Delta_2=25/T$, $\Delta_3=45/T$, satisfying Eq.~\ref{eq:delta1sum}. In that case, ${\mathbf{p}}_{\textrm{est}}=[0.30,0.25,0.45]$.
\end{Example}

\subsection{Case 1: A {{prior}} on onsets is available}

Let's first assume that the end-user can provide the expected onsets (composed of $K$ values). {{These}} {{prior}} onsets can, for example, be provided by a physics-based model able to predict the onset of each damage or cluster. Prior onsets can be represented as a distribution, say ${\mathbf{p}}_{\textrm{true}}$, following the same reasoning as before. The problem is now to compare ${\mathbf{p}}_{\textrm{true}}$ (the {{prior}} distribution) and ${\mathbf{p}}_{\textrm{est}}$ (the estimated distribution).

The degree of agreement between estimated onsets and the ground truth as quantified by ${\mathbf{p}}_{\textrm{est}}$ and ${\mathbf{p}}_{\textrm{true}}$ respectively, can be evaluated by different means, for example using the L$_1$ norm, the Hellinger distance, the Bhattacharyya distance, the total variation distance between probabilities, the Kullback-Leibler (KL) or Renyi divergences, among others. The KL divergence is used in the present work. KL is \textit{not symmetric} and \textit{never negative} and was used in many machine learning algorithms based on probability theory and on artificial intelligence. In information theory, KL is the amount of information lost when one distribution (for example, depending on parameters) is used to approximate another (for example, related to a ground truth). We compute the KL divergence as follows:
\begin{equation}
KL({\mathbf{p}}_{\textrm{est}} || {\mathbf{p}}_{\textrm{true}} ) = \sum_{k=1}^K {\mathbf{p}}_{\textrm{est}}(k) \log_2 \left(\frac{{\mathbf{p}}_{\textrm{est}}(k)}{{\mathbf{p}}_{\textrm{true}}(k)}\right)
\label{eq:kl1}
\end{equation}
In this expression, ${\mathbf{p}}_{\textrm{est}}$ is dependent on $S$, the subset of features used for clustering, and $K$ the number of clusters. The KL thus comprises two terms. First, a cross entropy loss 
\begin{equation}
CE({\mathbf{p}}_{\textrm{est}} || {\mathbf{p}}_{\textrm{true}} ) = -\sum_{k=1}^K {\mathbf{p}}_{\textrm{est}}(k) \log_2 \left({\mathbf{p}}_{\textrm{true}}(k)\right)
\label{eq:CEnt}
\end{equation}
 and second, Shannon's entropy
\begin{equation}
E({\mathbf{p}}_{\textrm{est}}) = -\sum_{k=1}^K {\mathbf{p}}_{\textrm{est}}(k) \log_2 \left({\mathbf{p}}_{\textrm{est}}(k)\right)
\label{eq:Entr}
\end{equation}
with $KL = CE - E$. 

\subsection{Case 2: Without a {{prior}} on onsets}

Even though some initial onsets could be obtained numerically (by simulations) or experimentally (for example using infrared thermography or strain sensors), it is more frequent, in practice, to be in the situation where there is no {{prior}} on onsets of clusters. Therefore, one approach to set ${\mathbf{p}}_{\textrm{true}}$ without {{prior}} is to make use of the \textit{principle of maximum entropy}, since ${\mathbf{p}}_{\textrm{true}}$ is a distribution. 
This ensures that unconscious arbitrary assumptions are not made, and the introduction of biases extrinsic to the data is avoided \cite{jaynes1957information,de2018introduction}. Therefore, \textit{in the absence of {{a prior}} on onsets, among all possible distributions of onsets for ${\mathbf{p}}_{\textrm{true}}$, we should select the one with the largest entropy that corresponds to the uniform distribution}.

When using the uniform distribution of onsets for ${\mathbf{p}}_{\textrm{true}}$ in CE (Eq.~\ref{eq:CEnt}), we obtain 
\begin{equation}
{\mathbf{p}}_{\textrm{true}} = \textrm{Uniform} \Rightarrow CE({\mathbf{p}}_{\textrm{est}} || {\mathbf{p}}_{\textrm{true}} ) = -\log_2\frac{1}{K} = \log_2{K}
\label{eq:CEntUnif}
\end{equation}
which means that KL tends to:
\begin{equation}
{\mathbf{p}}_{\textrm{true}} = \textrm{Uniform} \Rightarrow KL({\mathbf{p}}_{\textrm{est}} || {\mathbf{p}}_{\textrm{true}} ) = \log_2{K} - E({\mathbf{p}}_{\textrm{est}})
\label{eq:KLUnif}
\end{equation}
We can simplify Eq.~\ref{eq:KLUnif} by dividing by $\log_2 K$, leading to
\begin{equation}
\frac{1}{\log_2 K} KL({\mathbf{p}}_{\textrm{est}} || {\mathbf{p}}_{\textrm{true}} ) = 1 - E_\textrm{norm}({\mathbf{p}}_{\textrm{est}}) 
\label{eq:KLUniform22}
\end{equation}
where $E_\textrm{norm} \in [0,1]$ is the {normalised} entropy of ${\mathbf{p}}_{\textrm{est}}$. Therefore, for a given $K$, minimising the initial criterion in Eq.~\ref{eq:kl1} divided by $\log_2 K$ is equivalent to maximising the entropy of onsets. 
A possible criterion to sort subsets is thus:
\begin{equation}
Q_\textrm{proposed}^\textrm{ONSETS}(S,K): \quad \textrm{Given } K, \textrm{ find } \argmax_{S} \big[ E_\textrm{norm} \left( {\mathbf{p}}_{\textrm{est}} \left( \cdot| S, K\right)\right) \big] 
\label{eq:kzdkzqkd}
\end{equation}
where it is made explicit that $E$ depends on $K$ (the number of clusters) and $S$ (subsets of features). Therefore, for a given $K$, the criterion is used to sort the subsets according to the distribution of the onsets. 

\medskip

\begin{Example}
Let us continue the previous example, with $T=100$ and $K=3$ we can define three onsets distributed uniformly in order to define ${\mathbf{p}}_{\textrm{true}}$, located at $t_1^\textrm{true}=0$, $t_2^\textrm{true}=T/3=33.33$, $t_3^\textrm{true}=2T/3=66.66$, with all $\Delta_k=33.33/T$. Therefore ${\mathbf{p}}_{\textrm{true}}={\mathbf{p}}_{\textrm{uniform}}=[0.33,0.33,0.33]$ (at two digits of precision, the sum must be one). Using ${\mathbf{p}}_{\textrm{est}}$ of the previous example, we get $KL = 0.0454714$ (Eq.~\ref{eq:kl1}) with $CE = 1.585$ and $E = 1.539$. The maximum of $E$ for $3$ clusters is $\log_2(3)=1.585$ therefore $E_\textrm{norm} = 1.539/1.585=0.971$.
\end{Example}

\medskip

\begin{Remark}
One important point to observe is that the proposed CVI only requires the partition, not the features, conversely to a shape-based CVI. Therefore, according to \cite{vendramin2010relative}, the proposed criterion belongs to the family of ``external'' CVIs whereas shape-based CVIs are referred to as ``internal''. However, the proposed criterion is a special case because external CVIs actually require a complete reference partition to be compared with, meaning that the ground truth must be known for all feature vectors, whereas the proposed CVI only requires the onsets of clusters. 
\end{Remark}

\medskip

{{
\begin{Remark}
Our criterion can be used in an online setting for SHM applications, provided $T$ is known. $T$ represents, for example, the maximum number of cycles for fatigue testing, or the maximum crack size. Since in an online setting, $K$ is assumed to evolve, we need to iteratively store the onsets. The list of onsets will be updated with a new value (appending a new element in a vector) when a new cluster is created. Then, Eq.~\ref{eq:delta1} and~\ref{eq:delta1sum} can be computed, followed by Eq. \ref{eq:kzdkzqkd} (for example, without {{prior}}).
\end{Remark}
}}

\subsection{Proposed algorithm}

{{The criterion presented in the previous section is implemented within a broader procedure that incorporates the steps of the feature-based clustering method outlined in Figure \ref{fig:zfezefzfz}. This procedure proceeds as follows.

First, a standard clustering method (in this case, K-means) is applied to the dataset, assuming a range for the number of clusters and possible feature subsets. Unlike other methods, for each partition an additional characteristic, called the cluster onset, is defined as the first occurrence of that cluster in the dataset. The criterion, given by Eq.~\ref{eq:Entr} or Eq.~\ref{eq:kzdkzqkd}, depending on whether prior information on the onsets is available, is then calculated for each partition. The best partitions are selected based on the maximisation of this criterion (\textit{e.g.}, the top percentile of the values of $Q_\textrm{proposed}^\textrm{ONSETS}(S,K)$).
}}

Finally, the onsets of each cluster in the selected partitions are extracted and compiled into a histogram representing the cumulative distribution of onsets across all selected partitions. Similar to interactive clustering \cite{bae2020interactive}, the end-user can analyse this histogram, and based on the positions and amplitudes of peaks, they can choose to either accept, reject, or further refine the results—potentially triggering a re-clustering process by using different features or alternative clustering methods.

An algorithm detailing this procedure is provided in Alg.~\ref{algo_3}.

\begin{algorithm}[ht]
\SetKwInput{Input}{Inputs}\SetKwInput{Output}{Outputs}
\Input{ \\
$\quad$ Matrix of features $\bY$ with size $N \times d$ \\
$\quad$ Range of number of clusters $[K_\textrm{min}, K_\textrm{max}]$ \\
$\quad$ Subsets of features $\mathcal{S}$ 
}
\Output{ \\
$\quad$ Criterion \\
$\quad$ Histogram of onsets\\
}

\BlankLine
\For{$K \in [K_\textrm{min}, K_\textrm{max}]$}{
	\For{$S \in \mathcal{S}$}{
		\emph{// Apply the clustering method and store the partition and parameters}\;
		$[P(S,K), \theta(S,K)] \leftarrow$ \textsf{Clustering($\bY(:,S)$, $K$)}\;


		\vspace{0.2cm}
		\emph{// Compute quality according to how onsets are distributed}\;
		$Q_\textrm{proposed}^\textrm{ONSETS}(S,K)$ $\leftarrow$ \textsf{Onsets\_Clusters\_Quality($P$)}\;

	}
	
	\vspace{0.2cm}
	\emph{// Sort and select a few subsets of subsets of features for each $K$ using Eq.~\ref{eq:kzdkzqkd}. The threshold can be adapted according to the number of subsets considered}\;
	$q_{0.99} = \textrm{\textsf{quantile} with probability } 0.99 \textrm{ of } Q_\textrm{proposed}^\textrm{ONSETS}(S,K)$\;
	$S_\textrm{sub}(K) = \{ S: Q_\textrm{proposed}^\textrm{ONSETS}(S,K) >= q_{0.99}\}$\;
		
		
	\vspace{0.2cm}	
	\emph{// Find onsets of the selected partitions}\;	
	$\textrm{Onsets}(K) = \textrm{Find\_onsets}(F(K))$\;
	
}
\emph{// Accumulate evidence from various numbers of clusters using histogram}\;
\emph{// The number of intervals in the histogram depends on the application}\;
$\textrm{Acc} = \textrm{Histogram}\left( \textrm{Onsets}(\forall K) \right)$\;
\vspace{0.2cm}	
\emph{// Present the histogram to the end-user who can then either reject, accept, or be ignorant of the result, triggering a re-clustering of the data \cite{bae2020interactive}.}\; 

\vspace{0.2cm}

\caption{{{Onsets-based algorithm using Eq.~\ref{eq:kzdkzqkd} by evidence accumulation. 
 \label{algo_3}}}}
\end{algorithm}

{{The interpretation of the histograms can vary across applications and specificity of the dataset. The location of the peaks can be helpful for the end-user to conclude that some damages potentially start or grow from these locations. It can also lead to the conclusion that the extracted features or the clustering method are meaningful or not. Examples are provided in the following sections.}}







\section{Experiments}
\label{sect5} 

Three applications have been considered. Table \ref{appList} presents the details of their parameterization, including the number of features per subset, the possible number of clusters, and the clustering method. For each application, the results of the proposed onset-based CVI (Alg.\ref{algo_3}) and the shape-based CVI \cite{sause12} (Alg. \ref{algo_1}) are presented.


\begin{table}[ht]
\caption{Applications used to illustrate the proposed criterion. GMM stands for Gaussian mixture model, GK for Gustafson-Kessel, EDM for electrical discharge machining.\label{appList}}
\small
\centering
\begin{tabular}{|c|c|c|c|c|c|c|}
\hline 
\multirow{2}{*}{Application} & \# sensors & Total \#  &  \# features & \multirow{2}{*}{\# clusters} & Ground & Clustering  \\
             &  used &        features   & per subset		  & 			& truth      &   method  \\	
\hline
Bolted metallic & 1 & 19 & 4 & $\{2,3,4 \dots 10\}$ & Yes (7 classes) & K-means \\
structure & & & & & &  \\
\hline 
EDM µ-drilling & 1 & 24 & 3 & $\{4,6,8,10\}$ & No  & GK \\
hard material & & & & & & \\
\hline 
Glued hybrid& 4 & 19 & 3 & $\{4,6,8,10\}$ & No  & GK \\
tubular structure & & & & & & \\
\hline
\end{tabular}
\end{table}

\subsection{Dataset $\#1$: Monitoring of a bolted metallic structure}

{{In this section, three state-of-the-art shape-based CVIs used in the AE community, namely the Silhouette,  Davies-Bouldin, and Calinski-Harabasz CVIs, are used to sort the partitions with the voting scheme \cite{sause12} and compared to the proposed onset-based CVI to provide a benchmark. It is important to note that these CVIs are adapted for hard clustering methods like K-means. Soft clustering techniques such as Gaussian mixture models (GMM) or fuzzy C-means (FCM), which assign probabilistic memberships, may not align well with these metrics. Therefore, K-means is used in both cases as well as because it is a widely adopted method in AE literature. }}

\subsubsection{Setup and data description}

\begin{figure}[htbp]
\centering
\includegraphics[width=0.30\columnwidth]{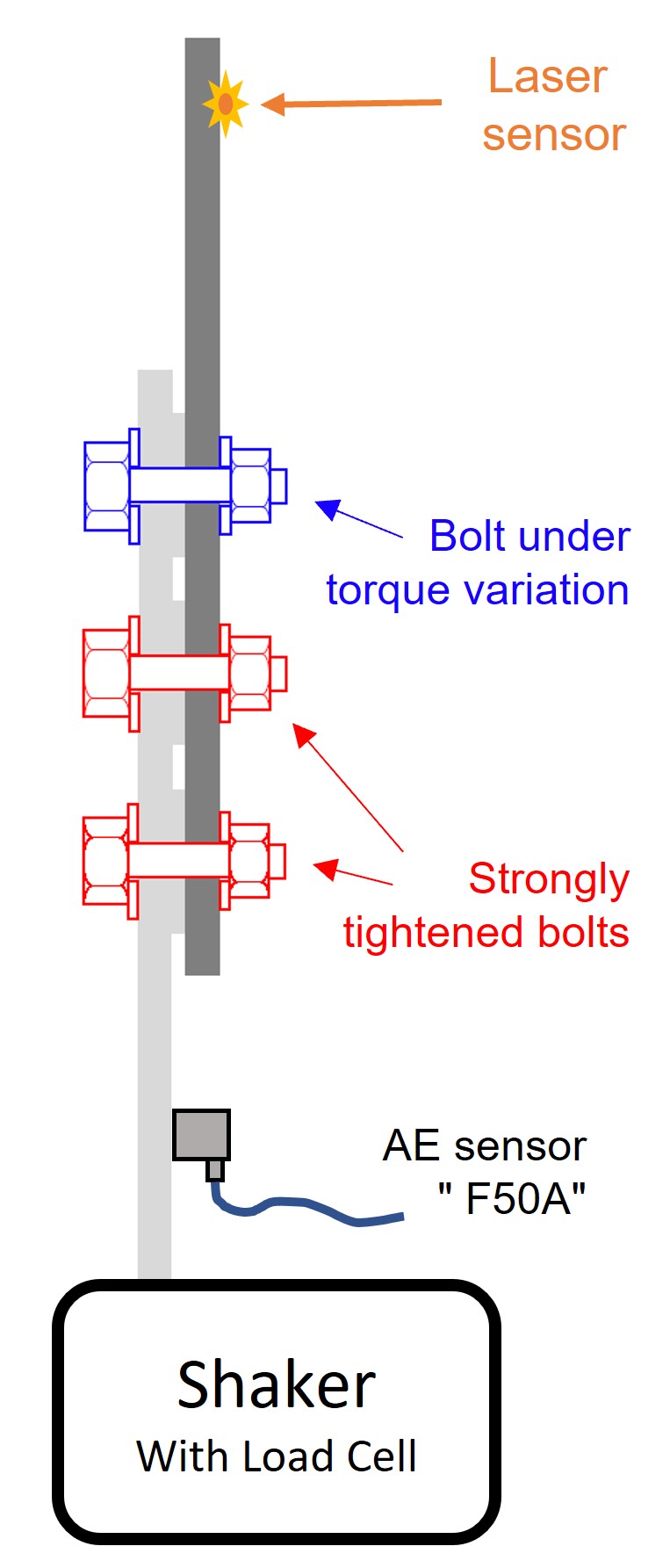}
\caption{Setup configuration for ORION-AE application. \label{fig:setup}}
\end{figure}

The benchmark dataset ORION-AE is presented in \cite{ORIONdata}. In this dataset, the true labels are provided, which allowed us to evaluate the performance of the proposed criterion and compare it with the standard solution. 
The data were collected on a setup designed to reproduce the loosening phenomenon observed in aeronautics, automotive, or civil engineering structures where parts are assembled together by means of bolted joints (Figure \ref{fig:setup}). 

\begin{figure}[ht]
\centering
\includegraphics[width=0.85\columnwidth]{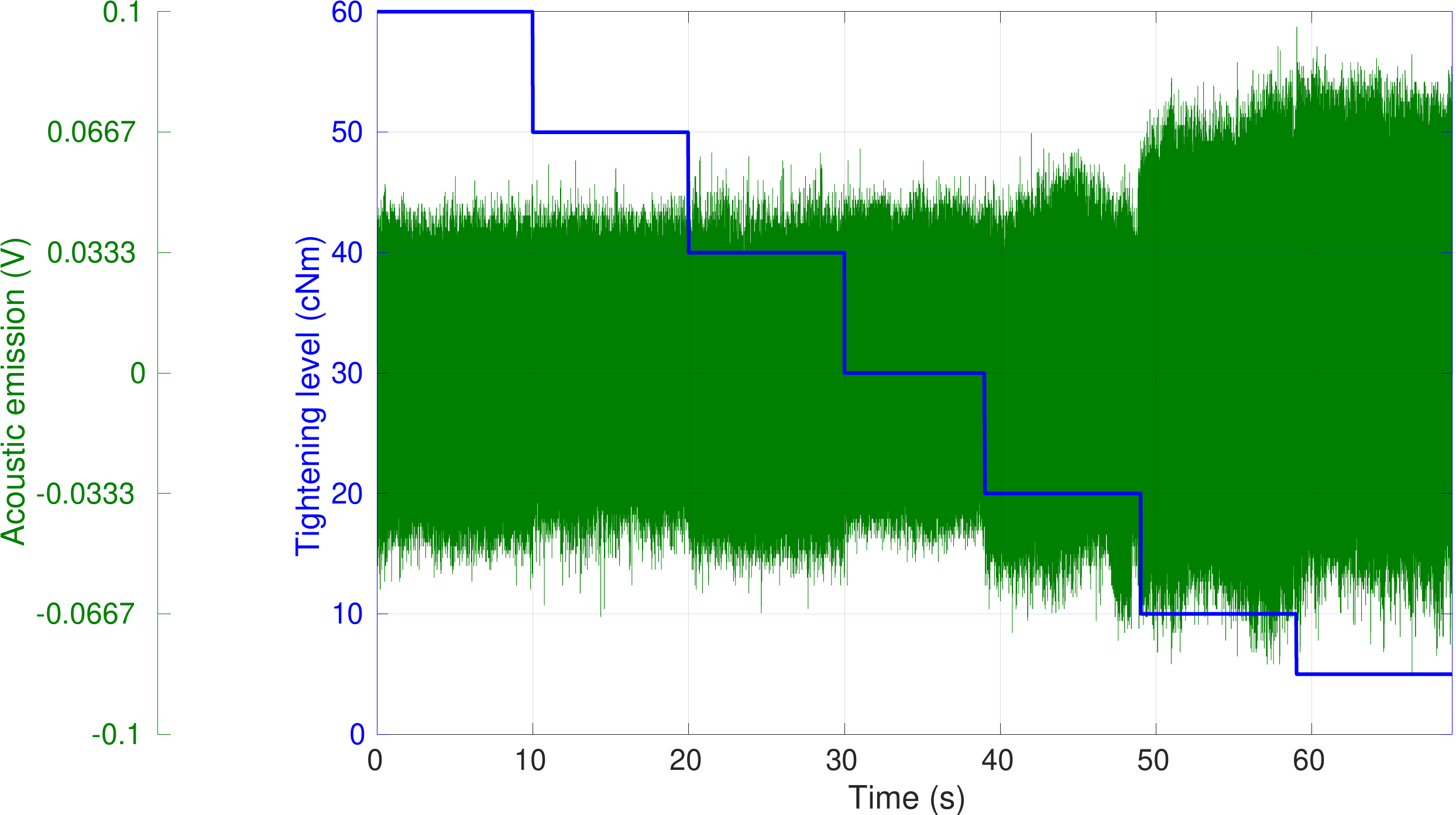}  
\caption{Raw data (green) and tightening levels (in blue represented by stairs).\label{rawdata}}
\end{figure}

The benchmark is composed of different campaigns of measurements. In the following, the set of measurements called ``$measurementSeries\_B$'' (campaign \#B) was used for illustration. The structure {{was}} submitted to a $120$ Hz harmonic excitation force. Seven tightening levels {{were}} applied on the upper bolt, starting from $60$ cNm (set with a torque screwdriver) and held during a $10$-seconds vibration test. The shaker {{was}} stopped and this vibration test {{was}} repeated after a torque modification at $50$ cNm. Torque modifications at $40$, $30$, $20$, $10$ and $5$ cNm {{were}} then applied in this order. In each campaign, three sensors are available and we used the AE sensor named ``F50A''  for illustration. The raw data are illustrated in Figure \ref{rawdata}. From the raw data (green), \textit{the goal is to infer the tightening levels by clustering and to compare the results with the ground truth} (blue stairs). 

The features are freely available at the following link \cite{ORIONdata}: \url{https://drive.google.com/drive/folders/1H413RxYu4ya7YMEgF_lTh_fHr7flvvOO?usp=sharing}. They are common features used in the AE community \cite{Kattis17}: Rise time, counts, counts to peak, duration, MARSE and absolute energies, amplitude, average signal level,  signal strength, RMS, reverberation frequency, initiation frequency, average frequency, partial power in [0,200], [200,400], [400,600], [600,1000] kHz, peak frequency, and frequency centroid. The set of feature vectors was post-filtered using a 31-sample moving median applied to each dimension in order to ensure temporal coherence, and one point in three were kept in order to reduce the amount of points (there are about 1200 cycles per tightening level in the original dataset).

\subsubsection{Parametrisation of algorithms}

Algorithms \ref{algo_3} and \ref{algo_1} were used on every combination of $4$ features within the set of $19$ mentioned above, resulting in $3876$ combinations for each $K=3,4\dots 10$ value with a total of $31008$ combinations. K-means clustering was run 5 times for every combination and the result leading to the minimum sum of squared Euclidean distances over all points and clusters was selected. If an empty cluster was obtained, the corresponding subset of features was not used. The set of partitions obtained was used in both criteria.
For Algorithm \ref{algo_1} (based on shape), the voting scheme proposed in \cite{sause12} was applied with three internal CVIs implemented in Matlab's Statistical and Machine Learning Toolbox: Silhouette, Calinski-Harabasz, and Davies-Bouldin. Algorithm \ref{algo_1} was applied for each $K$ independently in order to sort the subsets of features and select the best one. 
For Algorithm \ref{algo_3} (onset-based), the $20$ best partitions were selected for each number of clusters using the onset-based CVI. 

\subsubsection{Link between clusters onsets and change in tightening levels}

We evaluated whether the clusters' onsets coincide with the changes of tightening levels. For that, we represented the distribution of the estimated onsets using histograms. This was done in all partitions found by the previous algorithms and for all numbers of clusters ($K=3\dots 10$). Histograms depict the ground truth with blue stairs and the vertical dotted lines represent the instant of the true changes in the tightening level. This ground truth is available in \cite{ORIONdata}. The stairs show the duration of each level, equal to $10.0, 10.0,  10.0, 9.0170, 10.0, 10.0$, and $10.0$ seconds for $60, 50, 40, 30, 20, 10$, and $5$ cNm, respectively. 

\begin{figure}[hbtp]
\centering
\subfloat[\centering Shape-based CVI with the best vote \label{figB142a}]{\includegraphics[width=0.95\columnwidth]{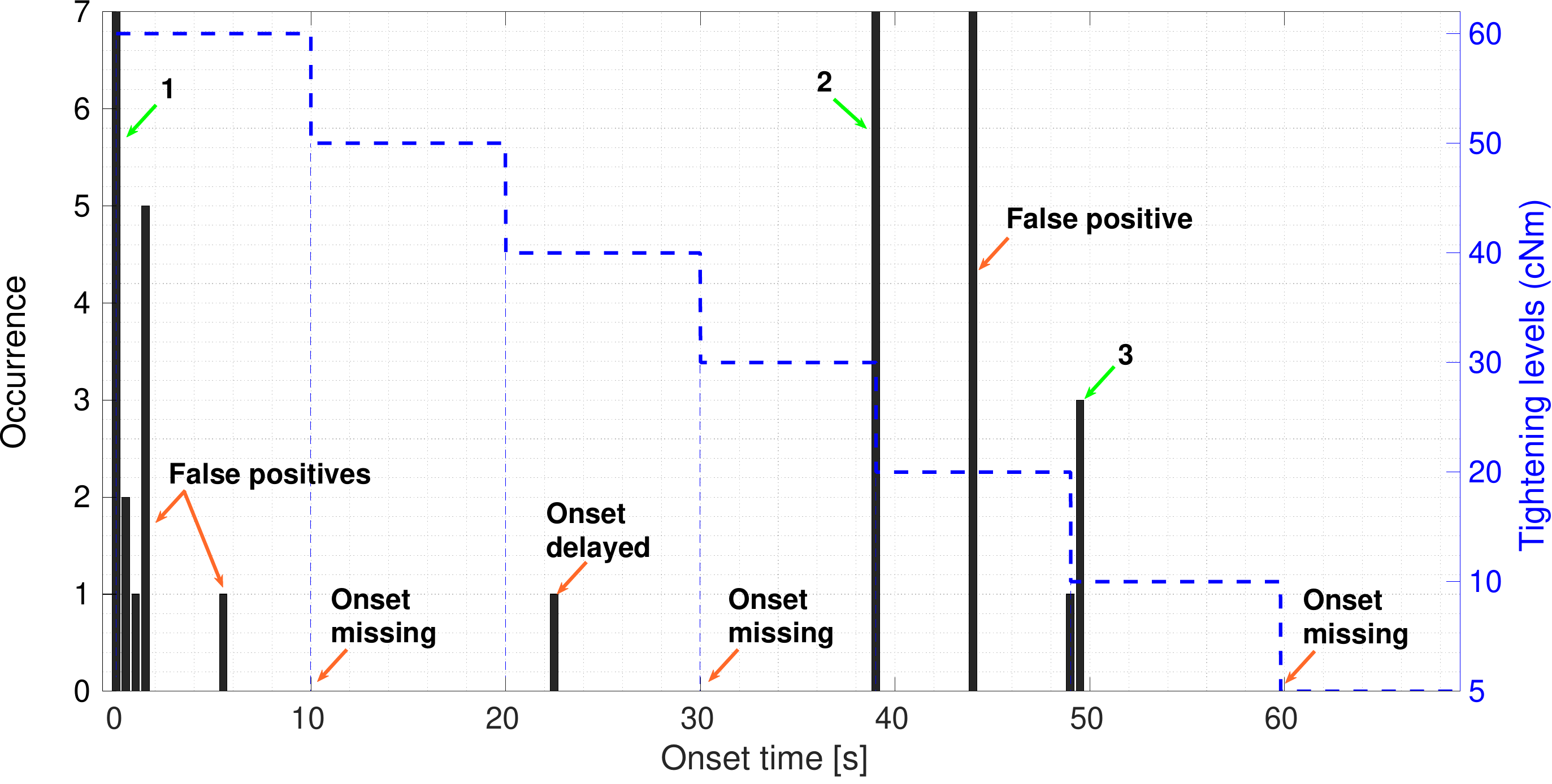}} \\
\subfloat[\centering Proposed onset-based CVI with 20 best partitions \label{figB142b}]{\includegraphics[width=0.95\columnwidth]{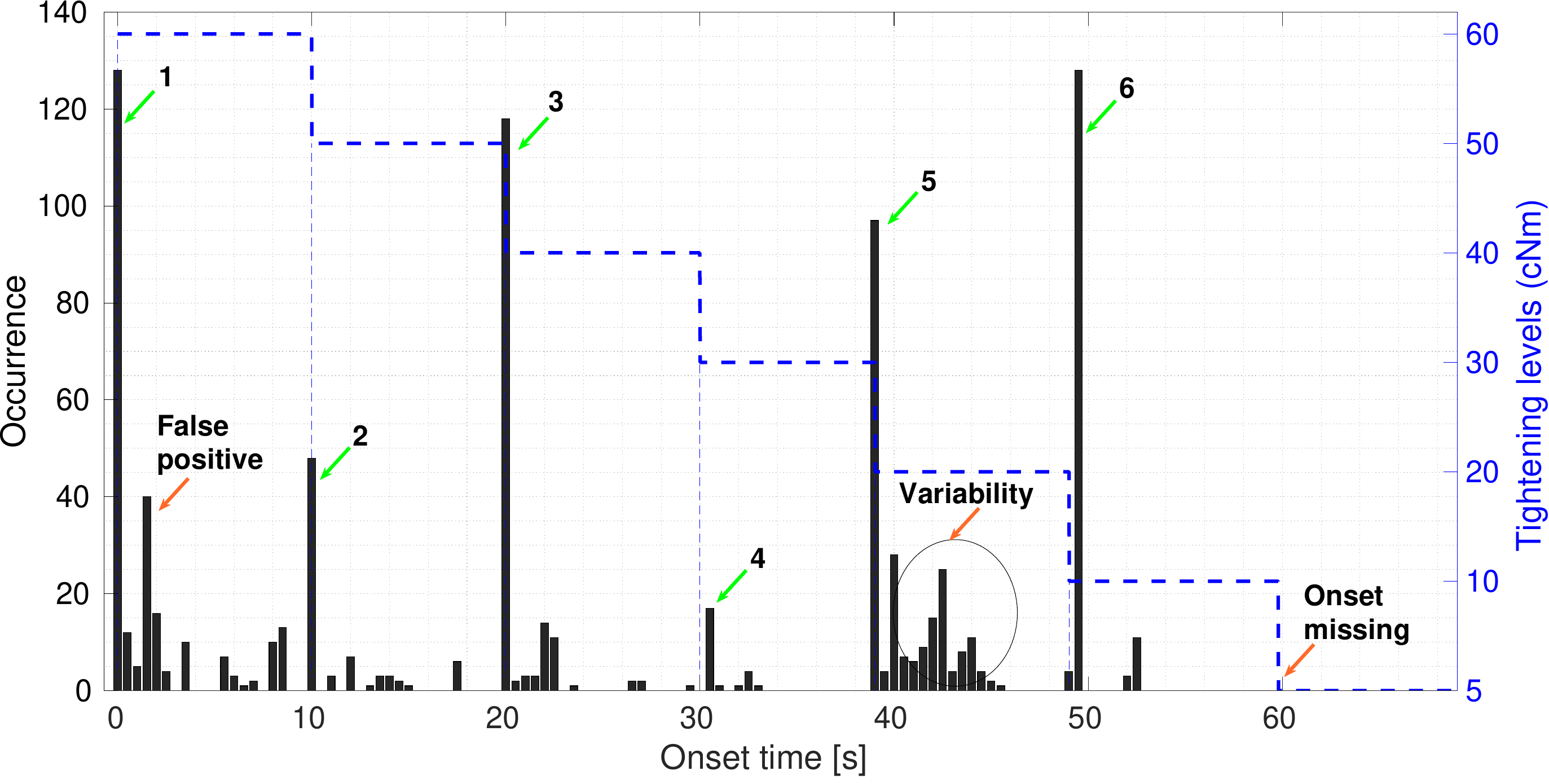}}
\caption{Validation of the clustering provided by a K-means using the voting scheme and onset-based criteria.\label{figB142}}
\label{fig:consseqB}  
\end{figure}

Histograms of onsets for shape-based and onset-based CVIs are presented in Figure \ref{figB142} with a bin size of $0.5$~s. \textit{These results only differ by the criterion since the same partitions obtained by the K-means method were used to evaluate both criteria in the same conditions}.
With the shape-based CVI (Fig.~\ref{figB142a}, we can observe three missing onsets (no vertical bar at 10, 30, and 60~s corresponding to 50, 30, and 5 cNm), one delayed onset (20 s / 40 cNm) and one clear false positive (at around 44~s). \textit{Based on this histogram, an end-user is likely to select 4 clusters, with one wrong and 3 missed tightening levels.}

In comparison, Figure \ref{figB142b} shows that all changes in tightening levels, except the last one, are detected by the onset-based CVI. Four primary peaks distinctly appear, located at expected times corresponding to changes in tightening levels at 60 cNm (0 s), 50 cNm (10 s), 40 cNm (20 s), 20 cNm (40 s), 10 cNm (50 s). \textit{The same partitions were used in both cases which means that only the CVI makes the difference. }The peak at 30~s is clearly distinguishable with 17 occurrences. The secondary peaks (lower occurrence) are globally distributed around the primary peaks and represent the variability on onsets in the 20 partitions.

{{High peaks mean that changing  the number of clusters and features does not modify the location of these onsets. However, as demonstrated by Sawan \textit{et al.} \cite{sawan2015unsupervised}, we expect that increasing the number of clusters gives different insights, which is clearly not the case with the shape-based CVI. 
Moreover, we observe that the number of small peaks for the onset-based CVI is larger, explained by the fact that 20 times more partitions were used, adding variability to the results. We observe that the average level of these peaks is below $10$, representing less than $8\%$ of values of highest peaks. This variability highlights the difficulty in detecting the tightening level at 40 cNm, which may be represented by more than one cluster. The high number of small peaks at the highest tightening levels (left-hand side) can be attributed to the challenge in accurately quantifying bolts loosening from vibration data \cite{niikura2021loosening,eraliev2022vibration,da2024domain}. 
}}

\subsubsection{Link between clusters onsets and clusters overall quality}

\begin{figure}[hbtp]
\centering
\subfloat[\centering Partitions selected by shape-based CVI]{\includegraphics[width=0.95\columnwidth]{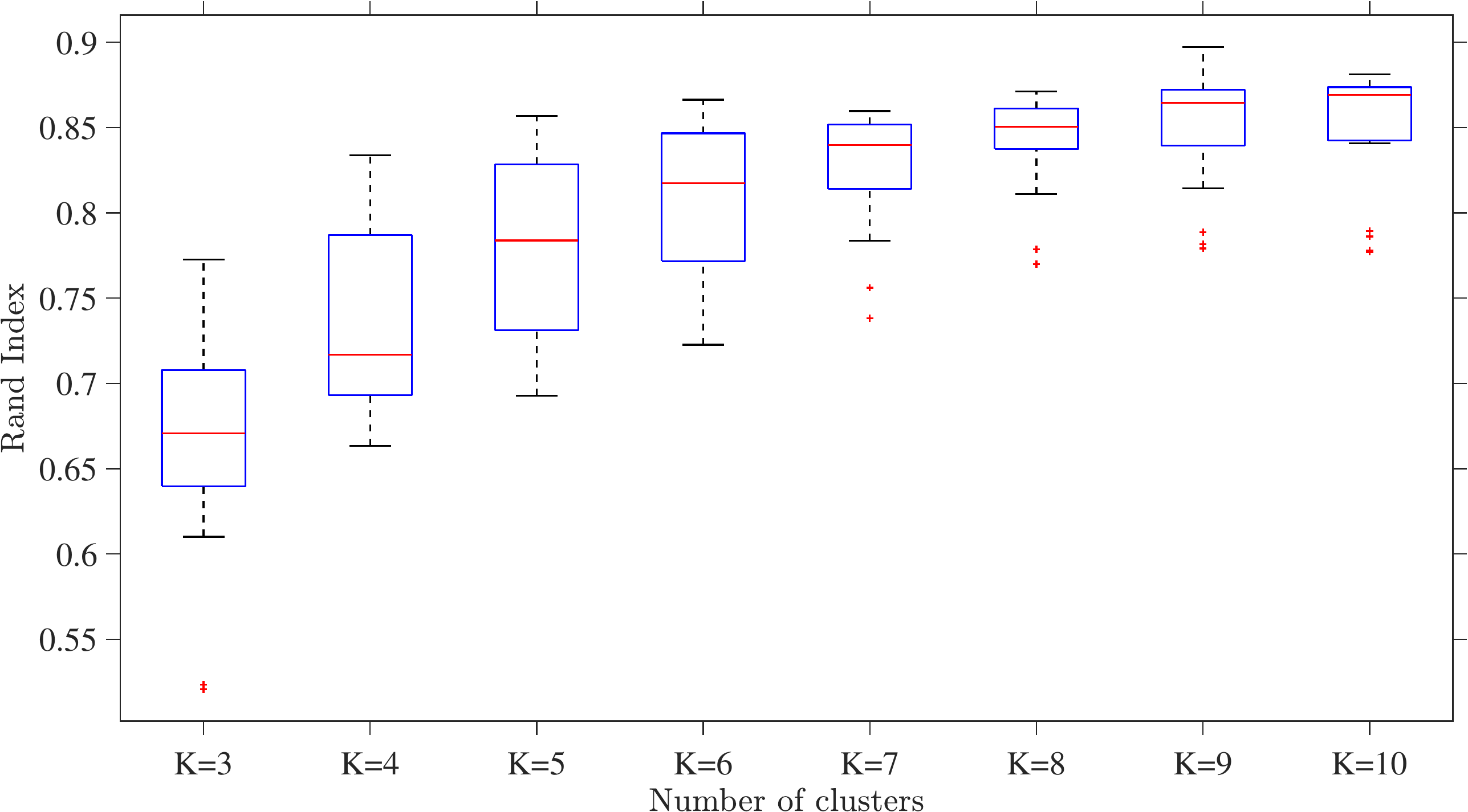}} \\
\subfloat[\centering Partitions selected by onset-based CVI]{\includegraphics[width=0.95\columnwidth]{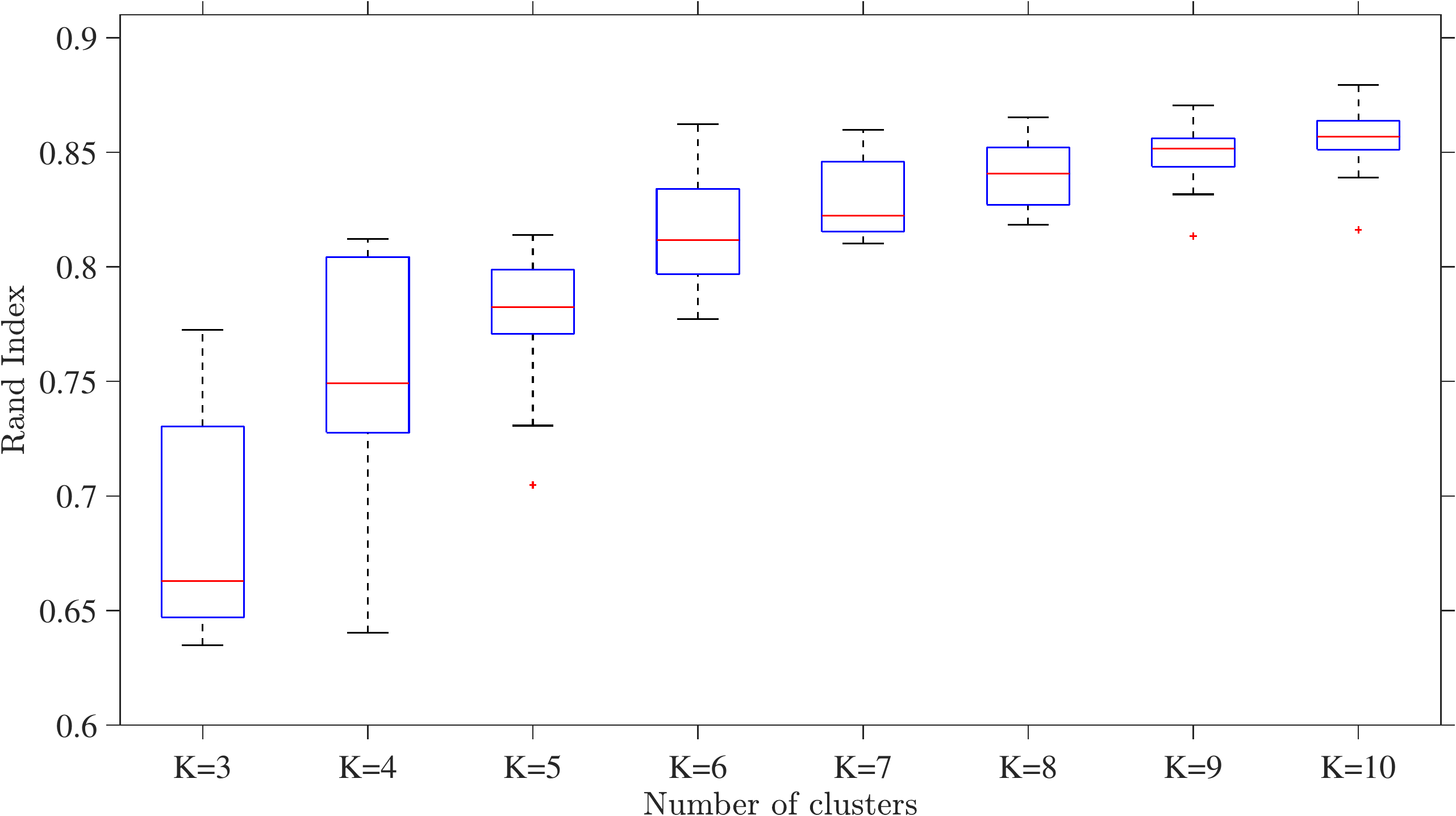}} 
\caption{Performance comparison evaluated by the Rand index. \label{figB142aaa}}
\end{figure}

{{
We have shown above that the onset-based CVI provides more relevant onsets. To further evaluate its performance, we assess whether the clusters generated coincide with the ground truth.}} For this, the 20 best partitions provided by each criterion were evaluated using the Rand index (RI) \cite{rand1971objective}. The RI is a value in the range $[0,1]$, where "$1$" indicates a perfect match between the clustering assignments and the ground truth. The number of points in each partition is 4684, based on which the RI is computed. {{The results are presented in Figure \ref{figB142aaa} using boxplots whereby the red line represents the median value, the boxes indicate variability based on the 25th and 75th percentiles, and the outliers are marked by red crosses.

We observe that the median RI values are generally close, indicating that, on average, both criteria provide similar cluster assignments across 4684 points. Note that the onsets are obscured in these results: only the histograms shown earlier allow for an assessment of their precise positions.

Concerning the variability across partitions, it differs between the two criteria. When considering values corresponding to the ground truth ($K_{true}=7$), the onset-based CVI demonstrates better consistency, with less variability in results compared to the shape-based CVI. This observation is confirmed in Table \ref{jdozdoza}, which summarizes the RI values for $K=6, 7, 8$. The minimum RI values show a 4\% to 7\% improvement for the onset-based CVI, while maximum values are comparable for both criteria. This suggests that the worst partitions identified by the onset-based criterion are more relevant than those produced by the shape-based CVI. Additionally, four outlier values below the minimum were found for the shape-based CVI (11 across all $K$ values), representing 10\% of the results. In contrast, no outliers were detected for the onset-based criterion at $K=6, 7, 8$, and only three for all $K$ values. These results demonstrate its superior robustness in response to feature changes. 
}}	

\begin{table}
\caption{Rand index for onset-based and shape-based CVI with K=6,7,8 ($\pm 1$ around the ground truth).\label{jdozdoza}}
\begin{tabular}{|c|c|c|c|c|c|c|c|c|}
\hline
\# cluster/perf& 	\multicolumn{2}{c|}{min} &	\multicolumn{2}{c|}{max}	& \multicolumn{2}{c|}{median} & \multicolumn{2}{c|}{\# outliers} \\
\hline
CVI				&	onset	& shape 		&	onset	& shape	&		onset &	shape &	onset &	shape \\
\hline
K=6				&	\textbf{\;77.7	}	& 72.3		& 	86.2		& \textbf{86.6}	&		81.2	  &	\,81.7  &	\textbf{0}	  &	0 \\		
\hline
K=7				& 	\textbf{\;81.0	}	& 73.8		&	86.0 	& 86.0 	&		82.2	  & \textbf{\;\;83.9	}  &	\textbf{0}	  &	2 \\
\hline
K=8				& \textbf{81.8}	& 77.0		&	86.5		& 87.1	&		84.1  & \textbf{\,85.9}	 & \textbf{0}	  &	2 \\
\hline
\end{tabular}	
\end{table}

\subsubsection{Case of a dataset with non-uniformly distributed onsets}

In the ORION-AE dataset, the tightening levels have approximately the same duration, which is a  special case. In this section, we propose to evaluate the proposed criterion when the duration in each tightening level differs. {{The dataset becomes imbalanced.}} For that, we selected only a part of the data in each level so that the duration of the levels follows a non linear evolution. Practically, it means that the metallic plates remain tightened for a longer period for a given level compared to the subsequent levels. For $60$, $50$, $40$, $30$, $20$, $10$ and $5$ cNm, the first seconds of data were kept as follows: $10\,s$, $8.5\,s$, $7\,s$, $5\,s$, $3\,s$, $2\,s$ and $1.5\,s$, and the remaining data were discarded.

\begin{figure}[hbtp]
\centering
\subfloat[\centering Shape-based CVI with the best vote]{\includegraphics[width=0.95\columnwidth]{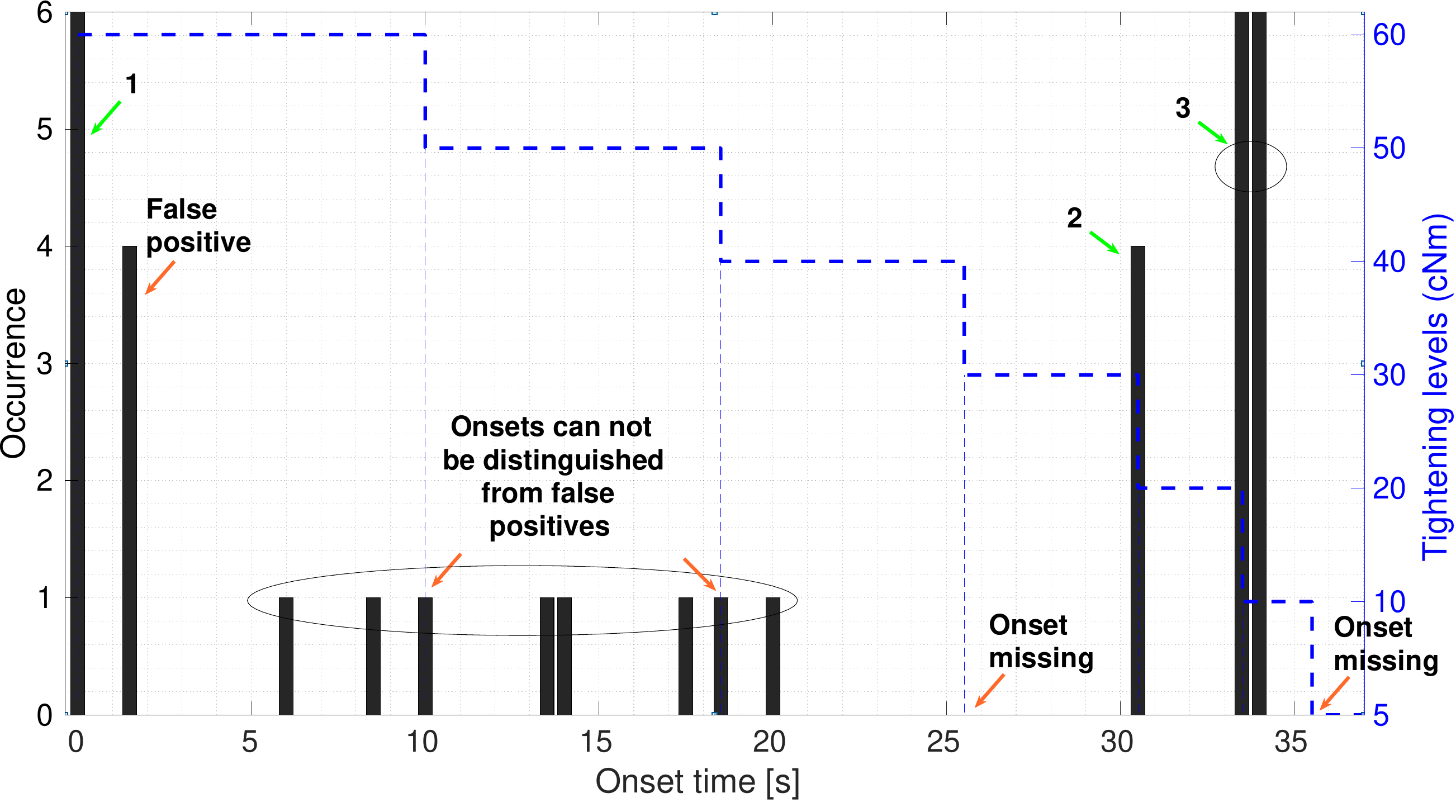}} \\
\subfloat[\centering Proposed onset-based CVI with 20 best partitions]{\includegraphics[width=0.95\columnwidth]{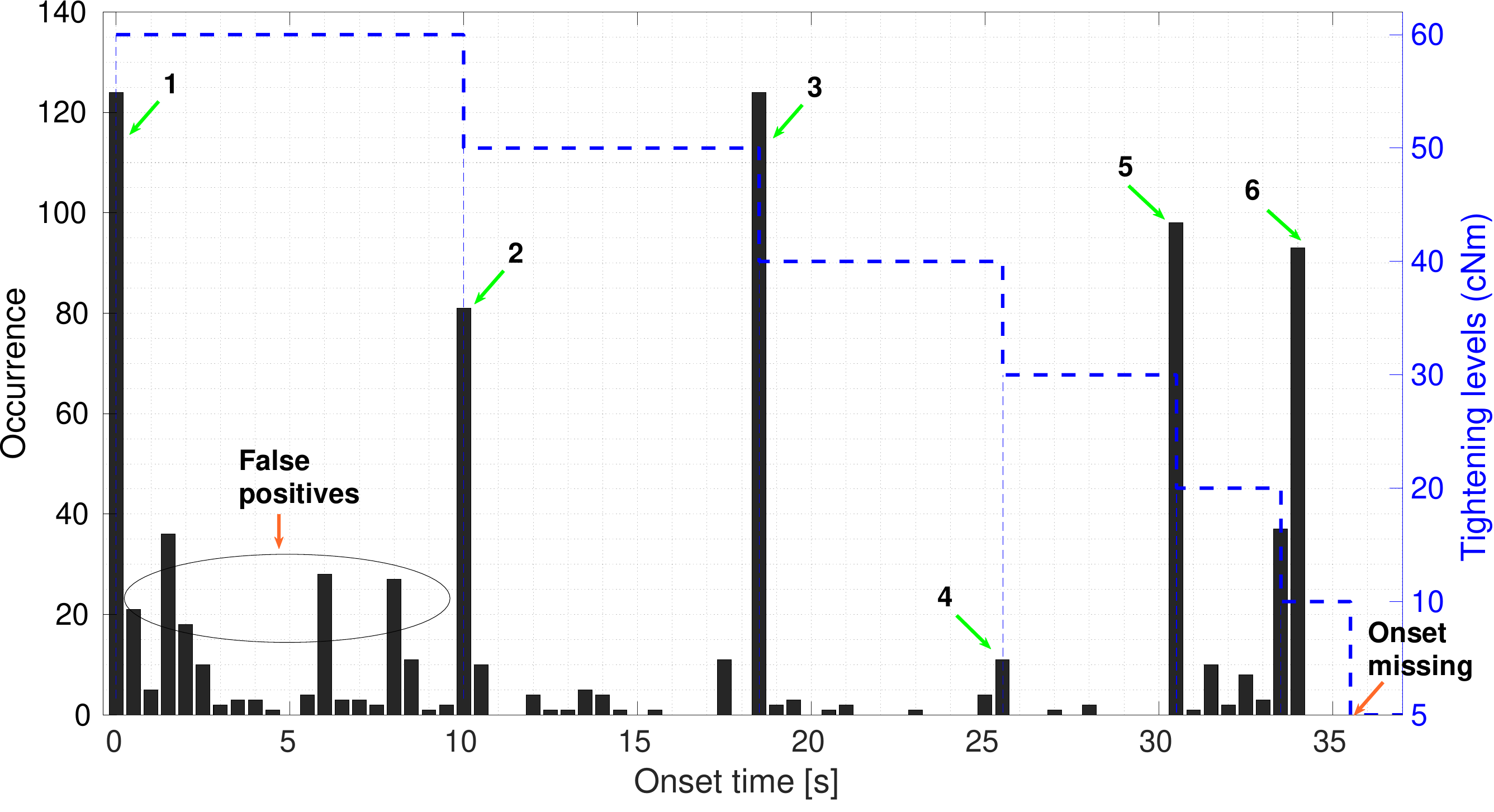}}
\caption{Case of non-uniform onsets: Validation of the clustering provided by a K-means using the voting scheme and onset-based criteria. \label{figB142dzdzd}}
\end{figure}

The results are reported in Figure \ref{figB142dzdzd}. We can observe that the proposed onset-based CVI outperforms the standard approach by selecting relevant partitions for which onsets are located as expected. As in the previous case, only the last onset is not discovered by the onset-based CVI. Conversely, the shape-based CVI only found three clear onsets. For 50 cNm (10 s) and 40 cNm (20 s) only one occurrence is found, which is not enough to distinguish these onsets from the false positives. 

{{The difference in behaviour of both criteria between this test (irregular timing of damage mechanisms) and the previous one (regular timing of damage mechanisms) is the size of the clusters. Since we shortened the duration of each tightening level, there are fewer points in the classes. Consequently, the distribution of the features within the classes has changed, making the clustering task more difficult. Some classes becomes smaller relatively to others due to this change of distribution which affects both the intra-cluster compactness and inter-cluster separability, therefore degrading the values of the shape-based CVIs. }}

\subsubsection{Summary}

When averaged over 20 partitions of \textasciitilde 4600 points, we observed that the quality of clusters found by the onset-based criteria is improved by $4\%$ to $7\%$, while using the same partitions as inputs, \textit{i.e.} all else being equal. With histograms, a distinct improvement was observed on the positioning of onsets with the proposed CVI, which makes this criterion suitable for structural health monitoring.

\subsection{Dataset $\#2$: Micro-drilling by electrical discharge machining}

This application is about monitoring of a manufacturing process known as micro-electrical discharge machining (micro-EDM) used for drilling  micro-diameter and deep holes. The goal of AE clustering is to gain insights about the process, and to evaluate if it is possible to detect when the electrode has reached the end of drilling and if the process proceeds similarly between two drilling events.

\subsubsection{Materials and method}
\begin{figure}[hbtp]
\centering
\subfloat[\centering \label{edm1}]{\includegraphics[width=0.45\columnwidth]{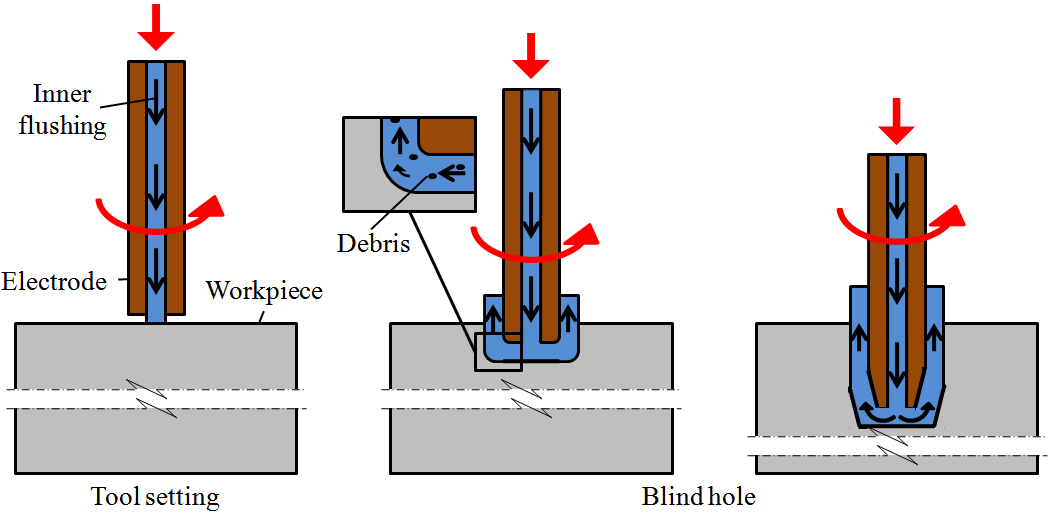}} \hspace{1cm}
\subfloat[\centering \label{edm2}]{\includegraphics[width=0.45\columnwidth]{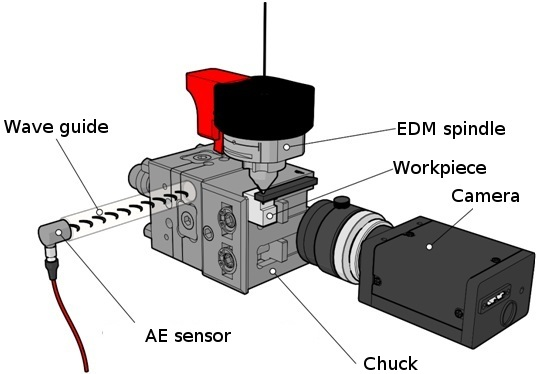}}
\caption{EDM principle (left) and instrumentation (right).\label{fig:envclust}}
\end{figure}

Micro-EDM relies on electrical discharges emitted by a tool-electrode to remove material from a work-piece immersed in a dielectric fluid (Fig.~\ref{edm1}). An electric arc forms between the conducting part and the electrode, which are not in physical contact, enabling the machining of very hard conductive materials with a high degree of flexibility in terms of shape. Both the feed rate of the tool-electrode and the sequence of discharges, which are necessary to remove material and achieve a specified depth, are controlled by proprietary software. During this process, both the work-piece and the tool-electrode experience wear, which is challenging and time-consuming to evaluate precisely and in real-time. Therefore, we used AE to collect data during drilling to gain insights into the process. Specifically, the goal is to assess whether the chronology of AE signals can aid in monitoring the progression of the electrode to the reference depth.

Experiments were conducted on a SARIX EDM machine, which features auto-regulation of input parameters for spark generation. A $0.3$ mm diameter tubular electrode was used to drill a $2$ mm hole into a $0.3$ mm thick cavity. The material being drilled is a Ni-based superalloy. Figure \ref{edm2} illustrates the setup on the EDM machine. The camera, synchronised with the start of the process, records the moment when the electrode begins drilling. It is then used to visually assess when a spark appears in the cavity, corresponding to the hole opening (third image in Figure~\ref{edm1}). This event is marked in the figures below as we anticipate specific behaviour in the AE signals and clusters.
AE data streams were acquired at a sampling frequency of 5 MHz using a "micro-80" sensor made by Mistras-EuroPhysical Acoustics, with an operating frequency range of 200-900 kHz. Data acquisition was performed using a Picoscope from Picotech. The AE hits were detected using the same method as in application \#1 (ORION-AE) and similar features, except for the partial powers, which were considered in the following intervals: [0 20], [20 100], [100 200], [200 300], [300 400], [400 500], [500 600], [600 800], and [800 1000] kHz, resulting in a total of 24 features.

\subsubsection{Results}

The tests described below were conducted using a brand new electrode and after cleaning the machine. Two tests were performed successively. Algorithm \ref{algo_3}, which is based on onsets, was applied to every combination of 3 features from a set of 24 (totalling 2024 combinations). The 20 best partitions for each number of clusters \( K \in \{4, 6, 8, 10\} \) were used to generate the histograms.

\begin{figure}[hbtp]
\centering
\subfloat[\centering Test \#1.\label{edmtest1}]{\includegraphics[width=0.95\columnwidth]{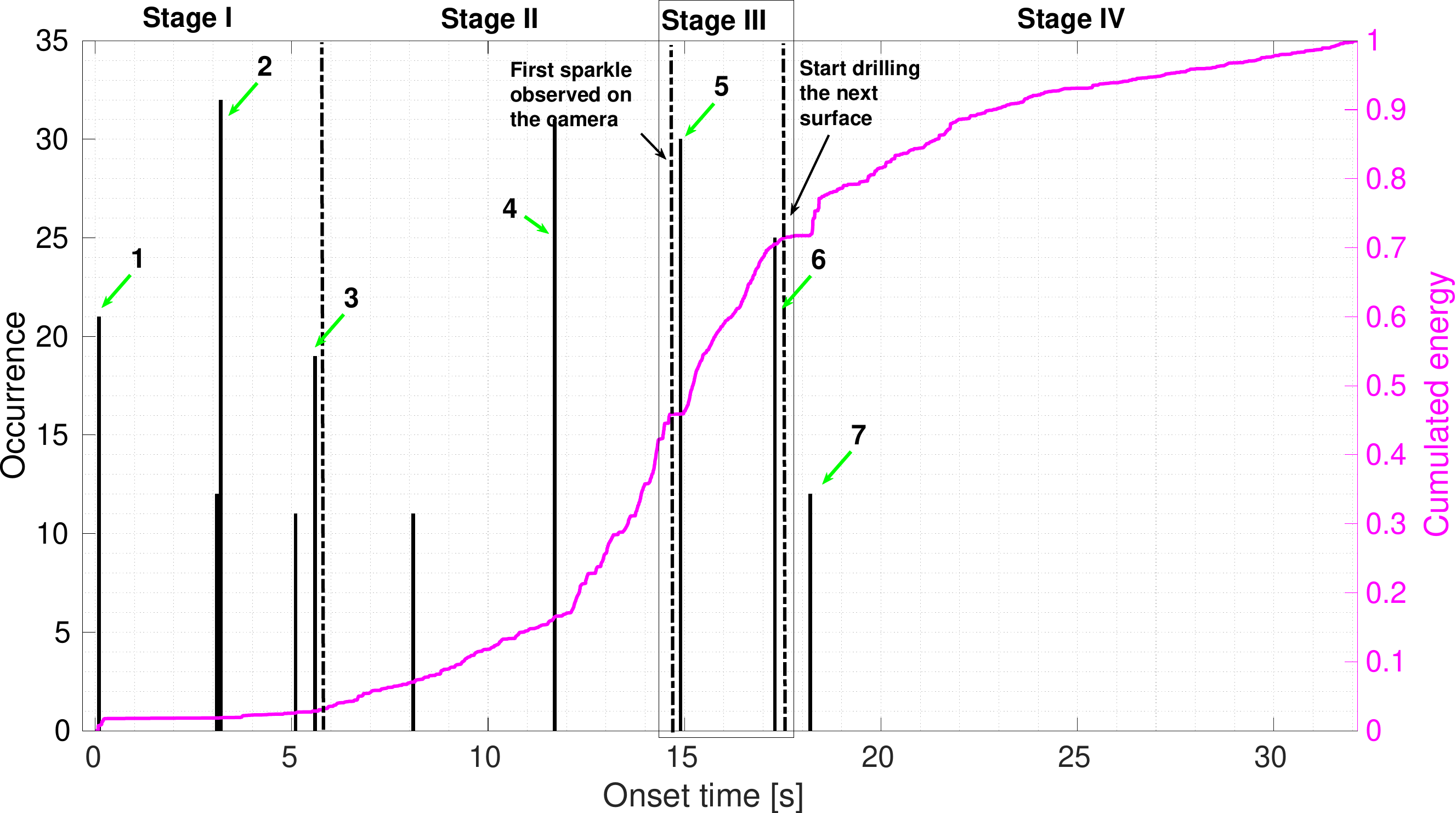}} \\
\subfloat[\centering Test \#2.\label{edmtest2}]{\includegraphics[width=0.95\columnwidth]{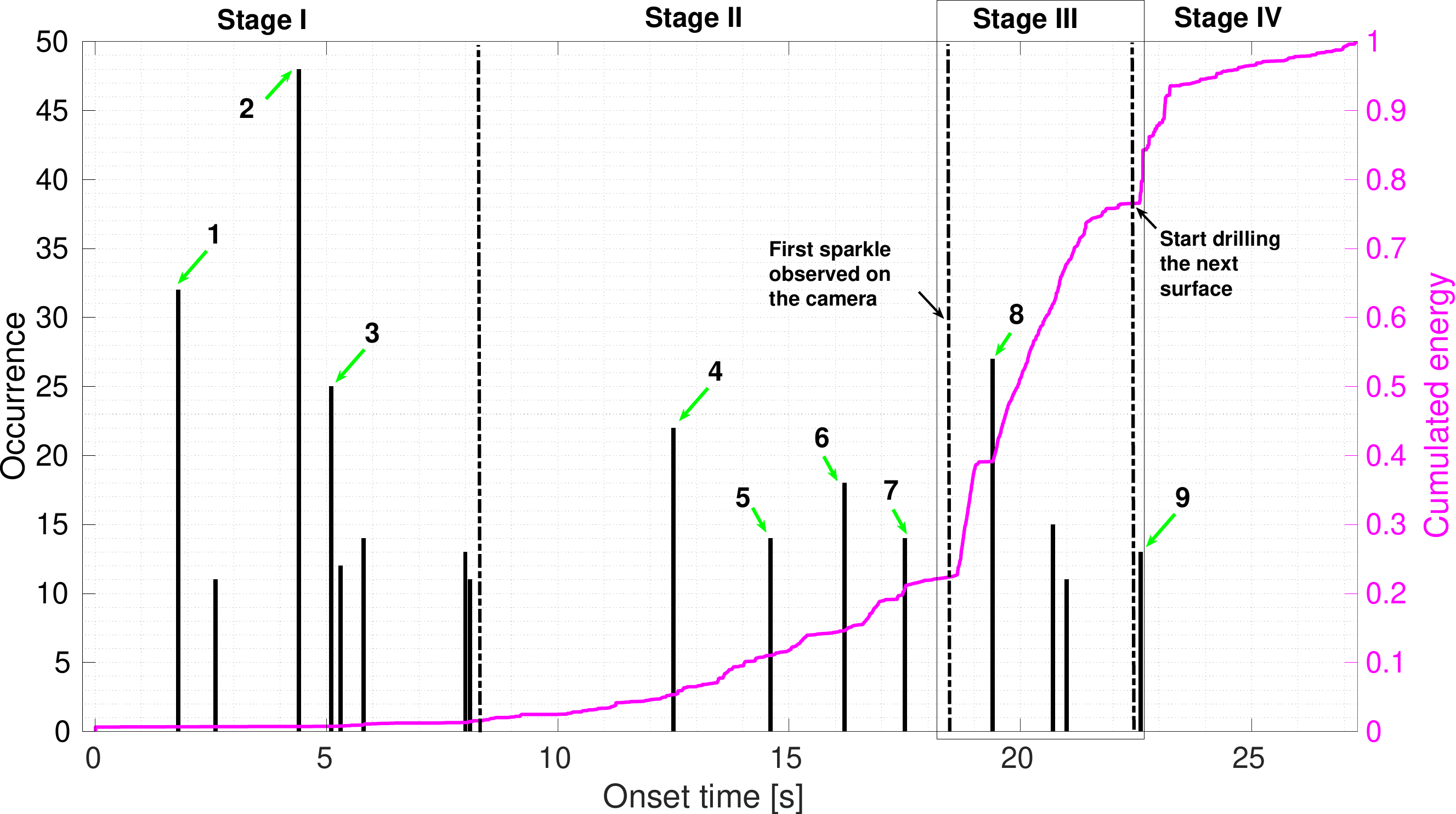}}
\caption{Histogram of onsets (peaks below 10 are not represented to lighten the figure). \label{edmtests12}}
\end{figure}

Figure \ref{edmtests12} presents histograms for Test \#1 (with a new electrode) and Test \#2 (after one use), showing peaks with more than 10 occurrences. Due to the continuous wear of the electrode during machining, its position was estimated using a camera to determine when the first spark appeared at the hole's opening. The time instances identified are: 1) Test \#1: First spark at 14-15 s; drilling from 17-18 s; 
Test \#2: First spark at 18-19 s; drilling from 22-23 s. Cumulated energy is superimposed on the histograms, as this feature traditionally indicates damage evolution trends.

For Test \#1 (Fig.~\ref{edmtest1}), four main stages are identified by the dashed lines, based on cumulated energy and onset positions. \textit{Stage I (0-6 s)} is characterised by three primary peaks with low energy, related to initial spark appearances. Drilling progresses with a few sparkles until a hole is made. \textit{Stage II (6-15 s)} shows significant increase in cumulated energy in two phases (6-12 s, 12-15 s), with a primary onset around 12 s indicating the machine's adaptation to drilling. \textit{Stage III (15-17.6 s)} depicts two primary onsets (5 and 6) and corresponds to the first spark observed by the camera, marking the approach to the first part of the work-piece. \textit{Stage IV (18 s onwards)} shows one more onset (7) indicating the start of drilling the second part, with similar characteristics to the first part.
For Test \#2 (Fig.~\ref{edmtest2}), Stage IV is shorter, and the process is more stable, starting at about 21.5 s. The cumulated energy increases more regularly, indicating a more stable drilling process. The rectangle highlights the first spark (around 18.2 s), with onsets before (7) and after (8) the observed spark. The end of this phase (just before starting drilling the next part) is marked by onset (9).

\subsection{Dataset \#3: Hybrid composite/metal joint structure}

This application is about monitoring of damages in a hybrid composite/metal joint structure. The proposed criterion is used to evaluate when damage mechanisms occur and how they grow during loading until the failure of the composite/metal joint. 

\subsubsection{Materials and method}

The specimen studied is a composite tube manufactured by Collins Aerospace, featuring a joint between a composite part and a metal part. The composite section consists of eight plies of filament-wound carbon fibre tows, forming a tube with a nominal internal diameter of 45.2 mm and a wall thickness of 4 mm. This geometry was chosen to handle loads typical of aerospace applications, where certifying bonded joints can be challenging.

The joint technology integrates a bespoke helical screw thread, where an epoxy-based adhesive is applied to the metal thread for lubrication during assembly. Preloading the joint applies through-thickness contact pressure to the composite, and loads are transmitted in two ways: (i) through reaction forces at multiple teeth and (ii) through friction at flat regions within the helical thread. If the frictional forces are exceeded, relative movement between the composite and metal parts can occur. Therefore, it is crucial to consider both slip and matrix damage in the design. The test was designed to ensure failure occurs at the interface regions rather than within the composite tube or metal end fitting.

\begin{figure}[htbp]
\centering
\includegraphics[width=0.75\columnwidth]{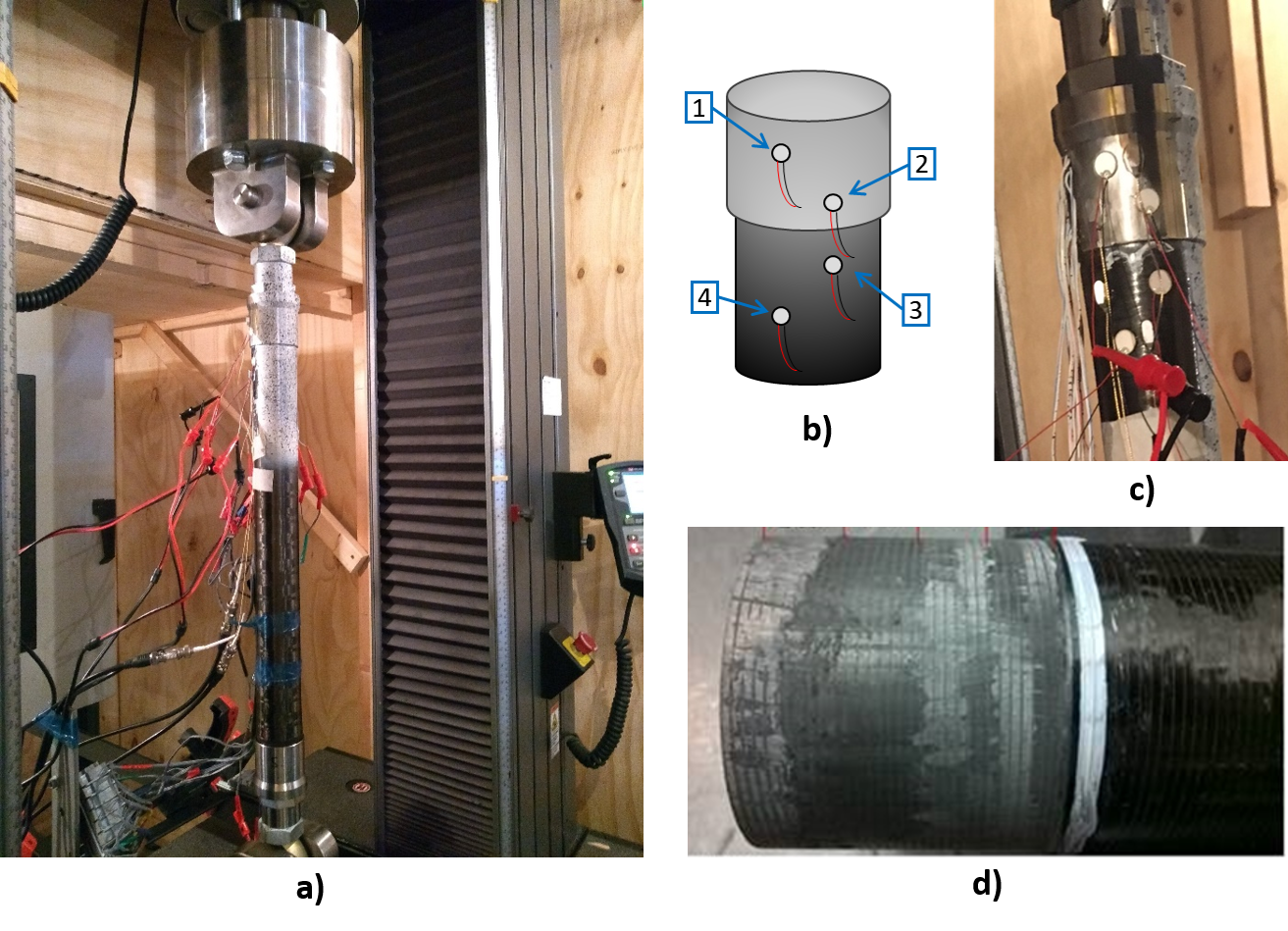} 
\caption{a) Quasi-static tensile loading of the hybrid composite/metal specimens. The specimen is connected to the loading rig at either end via a rod end with spherical bearings that are screwed into the metallic end fittings of the joint. b) and c) Schematic and photograph showing the arrangement of PWAS on both specimens. d) Photograph illustrating final failure of a hybrid composite/metal joint in the form of resin cracking and extensive slippage due to thread crushing. The joint was subjected to quasi-static tensile loading.  \label{setupneha}}  
\end{figure}

The specimen was tested in a quasi-static tensile loading configuration on an Instron 5989 electro-mechanical testing machine fitted with a 600 kN load cell. It comprises a composite tube and two metallic end fittings that form two joints, connected to the loading rig via rod ends with spherical bearings, which are screwed into the metallic end fittings. These rod ends are pinned to clevises that are screwed into the rig. The described test set-up is shown in Figure \ref{setupneha}-a).

The specimen was loaded at 0.5 mm/min crosshead speed in a single cycle until failure. The specimen failed at the instrumented joint. The breaking load was 268 kN. The final failure consisted of significant damage caused by initial matrix cracking and followed by slippage at the composite/metal interface of one of the two joints (Figure \ref{setupneha}-d) where the screw threads have been almost erased during failure. %

Piezoelectric wafer active sensors (PWAS) \cite{giurgiutiu2007structural} were bonded to the metal and composite parts of the joint for AE monitoring when the specimen is subjected to quasi-static tensile loading. PWAS of type PIC255, supplied by PI Ceramic-Germany, with 10 mm diameter and 0.5 mm thickness – were used. The positioning of PWAS is shown in Figure \ref{setupneha}-b) and~c). 

AE data were recorded by a PCI-2 based system, and captured in the software ``AEWin'' (Mistras, USA), with a sampling rate of 10 MHz and 20 dB of pre-amplification per sensor. Hit detection was performed by the software using a threshold of 55~dB and HDT=800~µs, PDT=200~µs and HDT=1~ms. In the following, only the data from sensors on the composite part (number 3, 4 on Fig. \ref{setupneha}b)) were used for the analysis. The same 19 features as in the first application were used. 

\subsubsection{Results}

{\it The objective of applying the onset-based CVI in this context is to assess its capability to reveal phenomena in a complex engineering structure and to detect incipient damages within the joint}. Data-driven methods like clustering can uncover patterns in large datasets, which is especially useful when prior knowledge is uncertain, as with the metal/composite joint scenario. This approach is beneficial when assigning labels is challenging and supervised algorithms are not feasible.

\begin{figure}[hbtp]
\centering
\subfloat[\centering \label{histoa}]{\includegraphics[width=0.95\columnwidth]{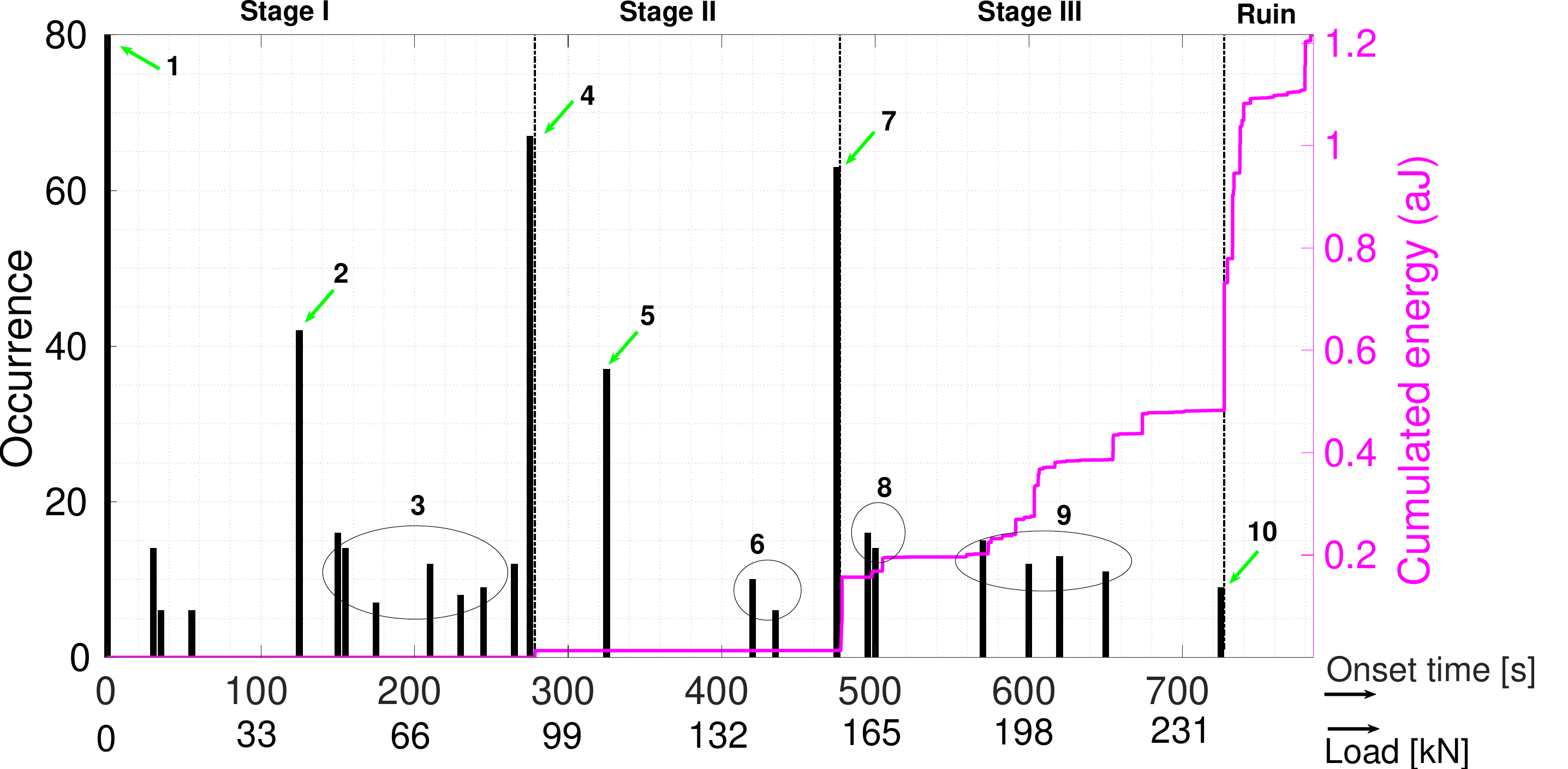}} \\
\subfloat[\centering \label{histob}]{\includegraphics[width=0.95\columnwidth]{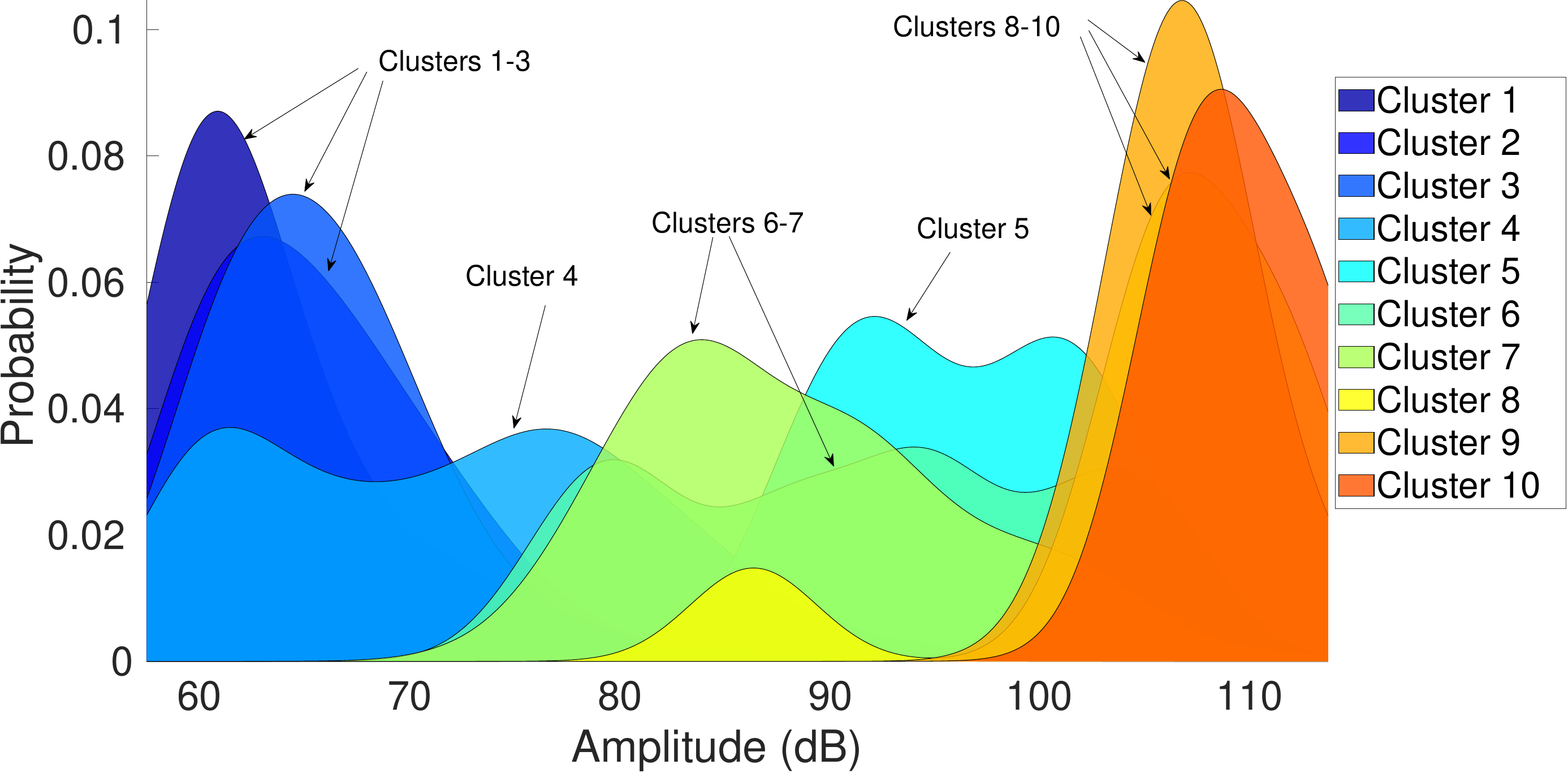}}
\caption{Analysis of clustering results for the third application. Left: Histogram of onsets according to time of test and load. Right: Distribution of amplitudes in clusters. \label{histo}}
\end{figure}

The Gustafson-Kessel clustering method \cite{gustafson1979fuzzy} was employed in Algorithm~\ref{algo_3}, considering all combinations of 3 features out of 19 (totalling 969 combinations) and varying the number of clusters $K \in \{4,6,8,10\}$, resulting in 3876 partitions. The 20 best partitions for each K were used to construct the histograms shown in Figure \ref{histoa}. Due to the lack of ground truth for this dataset, we compared the results with the cumulated energy from the previous application and provided the distribution of amplitudes in clusters for $K=10$ (Fig.~\ref{histob}).

The analysis reveals three distinct stages before failure. Each stage shows a unique evolution of cumulated energy and amplitude distributions. In the first two stages (0-275 s and 265-465 s), the cumulated energy exhibits two plateaus, indicating lower energy AE signals compared to the later stage III. This is confirmed by Fig.~\ref{histob}, where clusters in stage I have the lowest amplitudes (55-75 dB), stage II shows middle-range amplitudes (75-100 dB), and stage III has the highest amplitudes ($>$100 dB).

The histogram of onsets (Fig.~\ref{histoa}) shows a primary peak at 270 s, closely aligning with the initial increase in cumulated energy, marking a significant release of AE signals. This peak corresponds to primary peaks \#4 and \#5 at 265 s and 305 s, indicating incipient damage. During the plateau phase (275-465 s), fewer onsets are observed, with a significant peak (\#7) at 465 s, aligning with the start of increased cumulated energy and subsequent damage cascade.

Simulation results using Abaqus \cite[Chap. 7]{NehaPhD} show slip accumulation between 90-180 kN, followed by fatigue until failure at 280 kN (compared to 262 kN experimentally). This correlates with Stage II in the AE data, involving clusters \#4-8, suggesting that damages are represented by multiple clusters. The high number of onsets before 265 s (90 kN) and the low cumulated energy during this period may indicate friction, while the critical load accumulation at 465 s (160 kN) aligns with the fatigue period found in simulations (Stage III). Clusters around 600 s (78\% of test duration, 205 kN) represent incipient critical damages leading to final failure detected with cluster \#10. 


\section{Conclusion and future work}

{{
This study introduces a novel approach for interpreting acoustic emission (AE) data through a new clustering validation index (CVI) that emphasises the distribution of cluster onsets over time, cycles, or load scales. Unlike traditional methods that prioritise compact and separable clusters, our approach focuses on onsets, which are crucial for identifying different defects, such as mechanical loosening, process stages, or composite damage. This shift in focus is particularly valuable in monitoring applications where the initiation of defects is a key characteristic. 

\medskip 

The results demonstrated the criterion's ability to select optimal clustering results without prior knowledge, and to incorporate prior knowledge when available, which underscores its uniqueness and applicability.

In the bolted-joint benchmark, histograms of cluster onsets revealed that the onset-based CVI detected six out of seven true loosening events (missing only the final one), whereas the shape-based scheme missed three and introduced false positives. Quantitatively, the onset-based CVI improved the worst-case Rand index by up to 7\% and eliminated all low-quality outliers. This demonstrates that anchoring validation on the temporal distribution of cluster onsets yields more robust and interpretable partitions for SHM applications.



Applying the onset criterion to micro-EDM drilling revealed four distinct operational stages corresponding to tool engagement, stable drilling, hole breakthrough, and secondary penetration. Four main stages were identified by the onset-based CVI. The data-driven insights point to the potential for real-time quality control and tool-wear monitoring in industrial EDM processes.

In the hybrid composite–metal joint tests, onset histograms and amplitude-cluster distributions exposed three damage phases under quasi-static tensile loading: an early friction-dominated slip, a fatigue-driven plateau and a rapid damage cascade leading to catastrophic failure. The primary onset matched simulation-predicted fatigue onset at ~280 kN, while a secondary peak marked the transition to rapid interface degradation. These results highlight the criterion’s ability to detect incipient interfacial slip and matrix cracking without prior labeling, offering a powerful tool for complex joint monitoring.

\medskip


By focusing on when clusters appear rather than solely on their geometric compactness, the onset-based CVI aligns more directly with physical damage progression, making it especially suited for progressive degradation monitoring. However, its performance depends on sufficient temporal resolution and a known or consistent test duration. Adaptive schemes for estimating the duration or windowed onset tracking could be developed. Moreover, small clusters (e.g., very early or late events) can introduce variability, as seen in low-occurrence peaks; integrating amplitude or energy weighting could mitigate this.

Building on the aforementioned findings, future work will (i) integrate the onset-based index into supervised prognostics frameworks to trigger real-time alerts for defect initiation, (ii) explore hybrid CVIs combining onset information with traditional shape metrics for richer interpretability, and (iii) develop physics-informed priors for $p_{\rm{true}}$ to guide clustering in domains where simulations can predict expected damage timelines. Finally, embedding onset tracking within streaming algorithms will extend applicability to continuous SHM deployments.

}}

\appendix
\section[\appendixname~\thesection]{Voting scheme algorithm}

%
%

\label{app1}

Algorithm \ref{algo_1} was proposed in \cite{sause12}.

\begin{algorithm}[hbtp]
\SetKwInOut{Input}{input}\SetKwInOut{Output}{output}
\textbf{Input:} $\quad\quad$ Matrix of features $\bY$ with size $N \times d$\;
\textbf{Input:} $\quad\quad$ The range of number of clusters $[K_\textrm{min}, K_\textrm{max}]$\;
\textbf{Input:} $\quad\quad$ The subsets of features $\mathcal{S}$\;
\textbf{Input:} $\quad\quad$ The set of weights for two voting schemes $W_1$, $W_2$\;
\textbf{Output:} $\quad$ \,Voted partition of the data $P^{*}$\;
\textbf{Output:} $\quad$ \,Clusters parameters $\theta^{*}$\;
\BlankLine
\For{$S \in \mathcal{S}$}{\label{forins}
	\For{$K \in [K_\textrm{min}, K_\textrm{max}]$}{
		\emph{// Apply the clustering method and store the partition and parameters}\;
		$[P(S,K), \theta(S,K)] \leftarrow$ \textsf{Clustering($\bY(:,S)$, $K$)}\;
		\vspace{0.2cm}
		\emph{// Quality can be evaluated with one or several indices}\;
		\emph{// Indices used for AE are mostly internal and focus on clusters shape}\;
		$Q_\textrm{literature}^\textrm{SHAPE}(S,K)$ $\leftarrow$ \textsf{\textbf{Internal}\_Shape\_Indices\_Quality($P$, $Y(:,S)$, $\theta$)}\;
	}
}
\emph{// For each subset, firstly find the clusters using a voting scheme}\;
\emph{// The vote uses weights set a priori by the end-user}\;
\For{$S \in \mathcal{S}$}{
	$\textrm{nb\_clusters}(S) \leftarrow \argmax_{k} \textrm{Vote}\left( Q(S,k), W_1 \right)$\;
}
\emph{// Then find the subset and the final number of clusters using a second voting scheme}\;
$S^{*} \leftarrow \argmax_{S} \textrm{Vote} \left( Q(S,\textrm{nb\_clusters}(S)), W_2 \right)$\;
$K^{*} \leftarrow \textrm{nb\_clusters}(S^{*})$\;
\vspace{0.2cm}
\emph{// Return results}\;
$P^{*} \leftarrow P(S^{*},K^{*})$\;
$\theta^{*} \leftarrow \theta(S^{*},K^{*})$\;
\vspace{0.2cm}

\caption{Evaluation of clustering methods: Shape-based  \textit{internal} indices with voting scheme used in the literature \cite{sause12}. The output partition is on selected by a vote from the subset of partitions generated by clustering.
 \label{algo_1}}
\end{algorithm}


\end{document}